\documentclass[a4paper,11pt]{article}
\usepackage{jcappub}
\usepackage[T1]{fontenc}
\usepackage{graphicx}
\usepackage{amsmath}
\usepackage{hyperref}
\usepackage{breqn}

\title{Diffusion of cosmic rays at EeV energies in inhomogeneous extragalactic magnetic fields}
\author{Rafael Alves Batista}
\author{and G{\"u}nter Sigl}
\affiliation{ II. Institut f{\"u}r Theoretische Physik, Universit{\"a}t Hamburg \\ Luruper Chaussee 149, 22761, Hamburg, Germany}
\emailAdd{rafael.alves.batista@desy.de}
\emailAdd{guenter.sigl@desy.de}
\keywords{extragalactic magnetic fields, cosmic ray theory, cosmic web}

\abstract{
Ultra-high energy cosmic rays can propagate diffusively in cosmic magnetic fields. When their propagation time is comparable to the age of the universe, a suppression in the flux relative to the case in the absence of magnetic fields will occur. In this work we find an approximate parametrization for this suppression for energies below $\sim$ Z EeV using several magnetic field distributions obtained from cosmological simulations of the magnetized cosmic web. We assume that the magnetic fields have a  Kolmogorov power spectrum with the field strengths distributed according to these simulations. We show that, if magnetic fields are coupled to the matter distribution, low field strengths will fill most of the volume, making the suppression milder compared to the case of a constant magnetic field with strength equal to the mean value of this distribution. We also derive upper limits for this suppression to occur for some models of extragalactic magnetic fields, as a function of the coherence length of these fields.
}

\begin{document}
\maketitle
\flushbottom


\section{Introduction}

The discovery of cosmic rays dates back to the early 1910s. It is remarkable that the cosmic ray spectrum spans almost twelve orders of magnitude, from $\sim 10^9$ eV to $\sim 10^{21}$. Today, more than one century after the first observations of these particles, there are several aspects not fully understood, especially in the ultra-high energy domain. Ultra-High Energy Cosmic Rays (UHECRs) are very energetic particles ($E\gtrsim 10^{18}$ eV) that propagate through the universe and reach Earth. The sources of these particles are not yet known, but it is believed that they are mostly extragalactic \cite{auger2013b}. The mass composition is also under debate. Data from the Pierre Auger Observatory favors a heavy composition at the highest energies \cite{auger2010}, whereas the High Resolution Fly's Eye (HiRes) \cite{hires2010} and Telescope Array (TA) \cite{tinyakov2014} collaborations report a dominant light component. Despite the discrepancies at the highest energies, both experiments show a predominant light component at energies $\sim$ EeV (1 EeV = 10$^{18}$ eV).

Some features in the cosmic ray spectrum are noticeable. One happens approximately at 3$\times$10$^{15}$ eV and is known as the ``first knee''. Another one of these features, at about 8$\times$10$^{16}$ eV, is the so-called ``second knee'', observed by the KASCADE-Grande experiment \cite{kascade2011}. These two features seem consistent with a light (proton-dominated) component  accelerated up to the first knee, and a heavy component accelerated up to the second knee, as a result of the well known Peters' cycle \cite{peters1961}, in which the maximum acceleration energy of an element is proportional to its charge $Z$. In that case, the second knee would indicate the end of the galactic spectrum and the emergence of another galactic component, or an extragalactic one. 

A third interesting feature of the cosmic ray spectrum is the ``ankle'', at $E\approx$7$\times$10$^{18}$ eV\footnote{For a review on recent developments on the research of the ankle in light of the latest measurements, see ref. \cite{deligny2014}.}. It has been first observed by Linsley \cite{linsley1963} more than half a century ago, but its interpretation is still a matter of debate. In the original paper Linsley mentioned the possibility of the ankle being a signature of the transition between galactic and extragalactic cosmic rays, idea which still persists today \cite{wibig2005,allard2005}. Another interpretation for this feature was put forward by Berezinsky {\it et al.} \cite{berezinsky2005} in the context of the so-called dip model. In this model the ankle is a signature of pair production of UHE protons when interacting with background photon fields. This later interpretation requires a predominantly protonic component up to the highest energies and is in tension with data from Auger \cite{auger2010}, but not with HiRes \cite{hires2010} and TA \cite{tinyakov2014}.

The interpretation of the region between the second knee and the ankle is particularly fuzzy. From the theoretical point of view  it is difficult to accelerate galactic cosmic rays up to 10$^{18}$ eV through standard shock acceleration mechanisms. On the other hand, if the second knee marks the end of the galactic cosmic ray spectrum, one would need a new class of sources accelerating light elements to fill the gap between second knee and ankle, and still be consistent with the measurements \cite{auger2010,hires2010,tinyakov2014,kascade2013b}.

Regardless of the energy where the transition from galactic to extragalactic cosmic rays takes place, be it the second knee or the ankle, there probably is an energy below which the extragalactic component vanishes. This can happen, for example,  if we consider a suppression of the flux at ``low energies'' ($E \lesssim$ 10$^{18}$ eV) due to magnetic horizon effects spawned by diffusion of particles in extragalactic magnetic fields \cite{lemoine2005,aloisio2005,kotera2008}. 

The last interesting feature noticeable in the all particle cosmic ray spectrum is the suppression of the flux around 5$\times$10$^{19}$ eV, observed by Auger \cite{auger2008} and HiRes \cite{hires2008}. This suppression can be the so-called Greisen-Zatsepin-Kuz'min \cite{greisen1966,zatsepin1966} (GZK) cutoff, due to the interaction of UHE protons with cosmic microwave background (CMB) photons ($p+\gamma_{CMB} \rightarrow \pi^0 + p$). Another possibility is that the end of the spectrum is due to the maximum acceleration of the sources \cite{aloisio2011}.

Charged cosmic rays are deflected by the pervasive magnetic fields, namely the galactic and extragalactic. If their scattering length is  larger than their distance to the observer the propagation will be ballistic. If this length is much smaller, these cosmic rays will spatially diffuse. The typical environment where diffusion takes place are magnetized plasmas.  Particles can be magnetically scattered in different regions such as voids, filaments and galaxy clusters. In these regions the diffusion coefficients are very likely different, and so are the magnetic field strengths. In clusters of galaxies typical magnetic field strengths are $\sim$ 10 $ \mu$G with coherence length of $\sim$ 10 kpc \cite{carilli2002}. In the case of filaments the picture is not so clear (for a review on this topic see ref. \cite{ryu2011}), and estimates for the strength in these regions vary, with upper limits of the order of $\sim$ 0.1 $\mu$G, and coherence length ranging between 1 Mpc and 10 Mpc \cite{ryu1998,xu2006}.  In general, the coherence lengths of extragalactic magnetic fields are not known, and lie in the range between $10^{-12}$ Mpc and $10^{3}$ Mpc \cite{neronov2010}. Recent estimations based on gamma ray induced electromagnetic cascades suggest coherences lengths between $\sim$ 10 kpc and 1 Mpc \cite{neronov2013}. 

Syrovatskii \cite{syrovatskii1959} presented a solution for the difffusive propagation of particles from a single steady source. This solution was later generalized by Berezinsky and Gazizov \cite{berezinsky2006a} for an expanding universe. It is expressed in terms of the so-called Syrovatskii variable, which depends on the diffusion coefficient, which is energy and possibly position dependent. 

Mollerach and Roulet \cite{mollerach2013} addressed the problem of magnetic diffusion of UHECRs by assuming a scenario with a Kolmogorov turbulent extragalactic magnetic field. However, as explained before, different regions of the universe have different magnetic field strengths, which is probably related to the matter density in this environment, so that the strength of the magnetic field in clusters of galaxies is expected to be higher than in the voids (if magnetic fields in the voids really exist). Here we extended the aforementioned work by using magnetic field models from cosmological simulations performed by various authors.

\section{The Cosmic Ray Spectrum}

In this section we describe the mathematical framework underlying the diffusion of cosmic rays in magnetic fields. We follow Berezinsky and Gazizov \cite{berezinsky2006a} for the solution of the diffusion equation in a universe in expansion.

Let $n(E,\vec{r},t$) be the number density of particles with energy $E$ in an expanding comoving volume of the universe, at position $\vec{r}$ and time $t$. Assume that the diffusion coefficient is denoted by $D(E,t)$, and that the source has a generation function $Q(E,t)$. The diffusion equation for a source located at a comoving distance $r_s$ from the observer can be written as \cite{berezinsky2006a}:
\begin{dmath}
 \frac{\partial}{\partial t} n(E,\vec{r},t) - b(E,t) \frac{\partial}{\partial E} n(E,\vec{r},t) + 3H(t) n(E,\vec{r},t) - n(E,\vec{r},t)\frac{\partial}{\partial E} b(E,t) - \frac{D(E,t)}{a^2(t)} \nabla^2 n(E,\vec{r},t) = \frac{Q (E,t)}{a^3(t)} \delta^3(\vec{r}-{\vec{r}_g}),
 \label{eq:diffeq}
\end{dmath}
where $b(E,t)=-dE/dt$ describes the total energy losses, and $a(z)=1/(1+z)$ is the scale factor as a function of redshift $z$ for a source with comoving coordinates $\vec{r}_g=\vec{r}-\vec{r}_s$ at a distance $\vec{r}_s$ from the observer, and $H(t)$ is the Hubble constant at a time $t$. 

From the general solution of equation \ref{eq:diffeq} in the spherically symmetric case, one can write the flux $j$ for a single source \cite{berezinsky2006a}:
\begin{equation}
j(E, B, r_g) = \frac{c}{4\pi} \int\limits_{0}^{z_{max}} dz \left| \frac{dt}{dz} \right| Q(E_g(E,z),z) \frac{\exp\left( -\frac{r_g^2}{4\lambda^2} \right)}{(4\pi\lambda^2)^{\frac{3}{2}}} \frac{dE_g}{dE}.
 \label{eq:specss1}
\end{equation}
Here we use the standard $\Lambda$CDM cosmology, in which the redshift evolution is given by
\begin{equation}
 \left| \frac{dt}{dz} \right| = \frac{1}{H_0(1+z)} \frac{1}{\sqrt{\Omega_m(1+z)^3+\Omega_\Lambda}},
\end{equation}
with $H_0\approx67.04$ $\mathrm{km/s/Mpc}$, $\Omega_m\approx0.3183$ the density of matter in the universe, encompassing both baryonic and dark matter, and $\Omega_\Lambda\approx0.6817$ is the cosmological constant, assuming a flat universe ($\Omega_{tot}=1$) \cite{planck2013}. The source term $Q(E_g(E,z),z)$ can be assumed, following ref. \cite{mollerach2013}, as
\begin{equation}
 Q(E,z)=\frac{\xi_Z f(z) E^{-\gamma}}{ \cosh\left(\frac{E}{E_{max}}\right)},
\end{equation}
with $\xi_Z$ being the contribution of the nucleus of atomic number $Z$, and $E$ its observed energy, $f(z)$ a function for the redshift evolution of the source emissivity, $E_{max}$ the cutoff energy, and $\gamma$ the spectral index of the source. $E_g$ and $E$ are related in the following way \cite{berezinsky1988}:
\begin{equation}
 \frac{dE_g}{dE} = (1+z) \exp\left( \int\limits_0^z dz' \left| \frac{dt'}{dz'} \right| \frac{\partial}{\partial E_g} b(E_g,z')  \right),
\end{equation}
with $E_g$ denoting the initial energy of the particle at redshift $z'$ ($0<z'<z$), if the observed energy at present time is $E$. 

The variable $\lambda=\lambda(E,z)$ is the Syrovatskii variable\footnote{Our definition of $\lambda$ differs by a square from ref. \cite{berezinsky2006a}, and follows ref. \cite{mollerach2013}. This way, $\lambda$ has dimension of length and can be translated into the magnetic horizon. In this work we define the magnetic horizon as the distance that a charged particle can travel within a time interval equal to the age of the universe.}, first introduced by Syrovatskii \cite{syrovatskii1959} to address the problem of the distribution of relativistic electrons in the galaxy. The generalization of the Syrovatskii solution for an expanding universe was given by Berezinsky and Gazizov \cite{berezinsky2006a}, and can be written as
\begin{equation}
 \lambda(E,z,B)=\sqrt{ \int\limits_0^z dz' \left| \frac{dt}{dz'} \right| \frac{D(E_g,z', B)}{a^2(z')}}.
 \label{eq:syro2}
\end{equation}

\begin{figure}[h!]
	\centering
	\includegraphics[width=0.83\columnwidth]{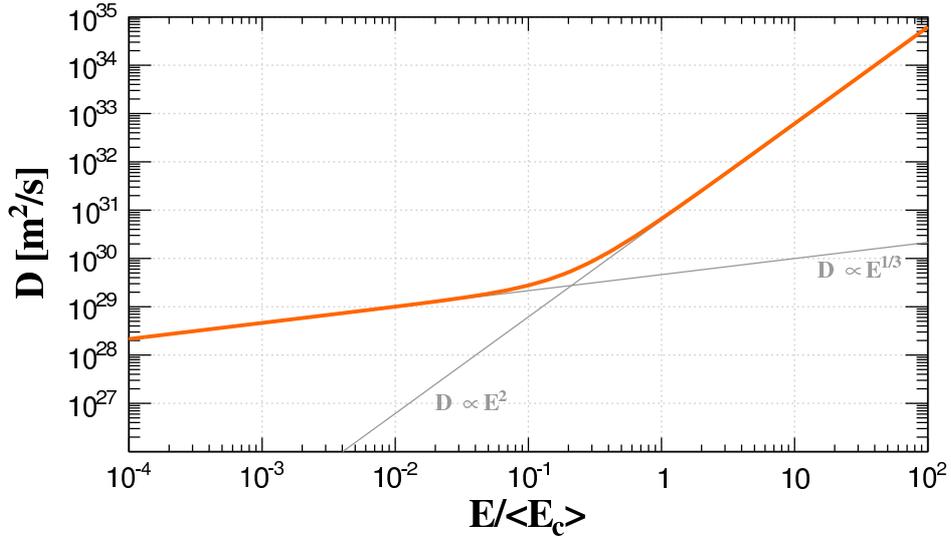}
	\caption{Diffusion coefficient as a function of $x\equiv E/\langle E_c\rangle$. Thin gray lines are the two energy dependent diffusion coefficients used to obtain the one adopted in this work, represented by the orange thick line. This case is for $z=0$.  }
	\label{fig:diffcoef}
\end{figure}

Following refs. \cite{aloisio2004,globus2008,mollerach2013}, we write the diffusion coefficient as a linear combination of the diffusion coefficients for the quasi-linear regime ($D \propto E^{1/3}$), dominant at lower energies, and the non resonant regime ($D \propto E^{2}$), dominant at higher energies:
\begin{equation}
    D(E, z, B) = \frac{cl_c(z)}{3} \left[  a_L \left( \frac{E}{E_c(z,B)} \right)^{\frac{1}{3}} + a_H \left( \frac{E}{E_c(z,B)} \right)^2 \right]   ,
    \label{eq:diffusion}
\end{equation}
which approximately holds for the resonant and non-resonant regimes, for the case of a Kolmogorov turbulence. Its behavior as a function of $x$ ($x\equiv E/\langle E_c\rangle$) is shown in figure \ref{fig:diffcoef}. The parameters $a_L$ and $a_H$ are, respectively, 0.3 and 4 \cite{mollerach2013}. Here $l_c(z)=l_{c,0}/(1+z)$ is the coherence length and $B(z)=B_{0} (1+z)^{2-m}$ the magnetic field strength as a function of redshift, with the subscript `$0$' corresponding to the value at present time, and $m$ a parameter due to the MHD amplification of the field. The critical energy $E_c=E_c(z,B)$ is defined as the energy for which the Larmor radius of the particle is equal to the coherence length of the fields, i.e., $R_L(E_c)=l_c$. It evolves with redshift as $E_c(z)=E_{c,0}(1+z)^{1-m}$. The Larmor radius is 
\begin{equation}
    R_L(E, B) = \frac{E}{cZeB} \approx \left( \frac{E}{\mathrm{EeV}} \right) \left( \frac{\mathrm{nG}}{B} \right){\ }\mathrm{Mpc} ,
    \label{eq:larmor}
\end{equation}
with $B$ being the magnetic field strength, and $E$ the energy. The explicit form of $E_c$ is, from equation \ref{eq:larmor}:
\begin{equation}
  E_c (z, B, l_c) = c Ze B(z) l_c (z) \approx 0.9 Z \left( \frac{B }{\mathrm{nG}} \right) \left( \frac{l_{c}}{\mathrm{Mpc}} \right){\  } \mathrm{ EeV},
  \label{eq:Ec}
\end{equation}
with $B=B(z)$ and $l_c=l_c(z)$.

The spectrum for a single source is shown in equation \ref{eq:specss1}. For $N_s$ sources, each one located at a distance $r_i$ from Earth, we can write the total spectrum as the sum of the spectra of individual sources, i.e.
\begin{equation}
 j_t(E,B) = \sum\limits_{i=1}^{N_s} j_s(E,B,r_i) = \frac{c}{4\pi} \int\limits_{0}^{z_{max}} dz \left| \frac{dt}{dz} \right| Q(E_g(E,z),z) \frac{dE_g}{dE} \sum\limits_{i=1}^{N_s}  \frac{\exp\left( -\frac{r_i^2}{4\lambda^2} \right)}{(4\pi\lambda^2)^{\frac{3}{2}}} .
 	\label{eq:spectot}
\end{equation}
Notice that if the number of sources is very large, the summation can be replaced by an integral. This integral can be calculated assuming spherical symmetry, yielding unity if the average separation between sources is small enough. Since in this case the dependence on the Syrovatskii variable will no longer be present, the spectrum will be independent of the modes of propagation and have the same shape regardless of the intervening magnetic fields. This result is known as the propagation theorem \cite{aloisio2004}. The spectrum obtained under these assumptions will be henceforth called universal spectrum.

We assume that the sources are located at finite distances from the observer, so we introduce a factor $F$, given by
\begin{equation}
	F = \frac{1}{N_s}  \sum\limits_{i=1}^{N_s} \frac{\exp\left( -\frac{r_i^2}{4\lambda^2} \right)}{(4\pi\lambda^2)^{\frac{3}{2}}},
	\label{eq:Ffactor}
\end{equation}
where the distances of the sources are defined according to ref. \cite{mollerach2013}:
\begin{equation}
	r_i = d_s \left( \frac{3}{4\pi} \right)^{\frac{1}{3}} \frac{\Gamma(i+1/3)}{(i-1)!}.
	\label{eq:ri}
\end{equation}
Here $d_s$ is the average separation between the sources, obtained from the source density.

Now we rewrite equation \ref{eq:spectot} as
\begin{equation}
 j_t(E,B)=\frac{c}{4\pi} \int\limits_{0}^{z_{max}} dz \left| \frac{dt}{dz} \right| Q(E_g(E,z),z) \frac{dE_g}{dE}  F(E, z, B).
\label{eq:spectot1}
\end{equation}
We can calculate the volume average of the spectrum by  weighting it by the magnetic field distribution, as follows:
\begin{dmath}
j(E) = \int\limits_{0}^\infty j_t(E,B) p(B) dB = \frac{c}{4\pi}  \int\limits_{0}^{z_{max}} dz \left| \frac{dt}{dz} \right| Q(E_g(E,z),z) \frac{dE_g}{dE} \left( \int\limits_{0}^\infty dB  F(E, z, B) p(B) \right),
\label{eq:spectot2}
\end{dmath}
where $p(B)$ is the probability distribution function corresponding to the magnetic field distribution obtained from the filling factors shown in figure \ref{fig:fillingFactors}.

We can only provide a volume averaged description of the spectrum. This is formally not the same as  taking the volume average of the diffusion coefficient, but the difference between these two approaches is small, particularly for higher redshifts ($z>$0.5), which correspond to most of the flux at $E\lesssim$ 1 EeV. A rough estimate of the relative difference ($\Delta F$) between the factors $F$, shown in equation \ref{eq:Ffactor}, obtained using these two approaches gives $\Delta F \lesssim $6\% at 10$^{16}$ eV and $\Delta F \lesssim $0.02\% at 10$^{18}$ eV. Notice that in this work we weight $F$ by the magnetic field distribution, as shown in equation \ref{eq:spectot2}.

The diffusion equation (equation \ref{eq:diffeq}) is solved assuming an energy and time dependent diffusion coefficient which, however, is constant in space.  When writing the diffusion equation\footnote{For a detailed derivation of the diffusion equation see appendix \ref{app:appendixA}.} one has the following term:
\begin{equation}
	\vec{\nabla} . \left[  D(\vec{x},t) \vec{\nabla} n(\vec{x},t) \right] = D(\vec{x},t) \nabla^2 n(\vec{x},t) +  \vec{\nabla}D(\vec{x},t) . \vec{\nabla}n(\vec{x},t).
	\label{eq:convterm}
\end{equation}
with $D$ being the diffusion coefficient, $n$ the number density of particles, $\vec{x}$ the physical coordinates and $t$ the time. If we assume $D$ is position independent, then only the first term in the right hand side of this equation remains, and $\vec{\nabla}n(\vec{x},t)$ can be neglected. This approximation is valid as long as $D(\vec{x},t) \nabla^2 n(\vec{x},t) \gg  \vec{\nabla}D(\vec{x},t) . \vec{\nabla}n(\vec{x},t)$.

The second term in the right hand side of equation \ref{eq:convterm} can be understood as a convection velocity, which means that the condition $D/\ell_D \ll c$ should hold, where $\ell_D$ is the length scale on which $D$ varies. This assures that the convection velocity $\sim D/\ell_D$ is much smaller than the speed of light everywhere. If the magnetic horizon effect is small, the average source distance will be of the order of a sizable fraction of the Hubble radius $R_H$. The above condition then implies that the neglected convective
term will only contribute for sources closer than $\sim(D/\ell_D c)R_H\ll R_H$ which give a subdominant contribution to the observed flux. Equivalently, this condition assures that the first term in eq. \ref{eq:convterm} indeed dominates over the second, assuming that the cosmic ray density varies on a length scale $\ell_D$ in regions of significant variation of $D$.

We will now consider for which range of parameters the above condition is satisfied, for energies $\sim$EeV. In the center of galaxy clusters we have $B\sim\mu$G and $l_c\sim$kpc. Therefore, from eq. \ref{eq:Ec}, the critical energy is $E_c\sim$EeV,  and using equation \ref{eq:diffusion} one obtains $D/c\sim$0.1 kpc. Thus, with $\ell_D\sim$100 kpc one has $D/\ell_D\sim$10$^{-3} c$. At the other extreme, in the voids, if $B\gtrsim0.1\,$ nG and $l_c\lesssim$10 Mpc, one has $E_c\gtrsim0.1(l_c/{\rm Mpc})$ EeV and thus $D\lesssim100(l_c/{\rm Mpc})^{-1}$ Mpc. Therefore, with $\ell_D\sim100$ Mpc on void scales one has $D/(c\ell_D)\lesssim$1 such that for $B\lesssim$0.01 nG and/or $l_c\lesssim$1 Mpc our approximation breaks down. In this case the spectrum may become somewhat flatter than predicted in our approach because there will be additional contribution to the flux from the neglected convection term, which depends on the energy.

It is important to mention that for the sake of numerical calculations we consider only adiabatic energy losses due to  the expansion of the universe. Pair production starts to become relevant for energies $\gtrsim$ 3Z EeV and pion production above $\gtrsim$ 50 EeV (for protons). Photodisintegration can also be neglected, for it conserves the Lorentz factor of the particles, hence keeping diffusion properties approximately unaltered. Therefore, it is a reasonable approximation to neglect all other energy loss processes if we are interested in energies $\lesssim$ Z EeV.

\section{Magnetic Fields from Cosmological Simulations}

In the present work we have considered the effects of extragalactic structured magnetic fields obtained from several cosmological magnetohydrodynamical (MHD) simulations, namely the ones performed by Miniati \cite{miniati2002}, Dolag {\it et al.} \cite{dolag2005}, Das {\it et al.} \cite{das2008} and Donnert {\it et al.} \cite{donnert2009}. Das {\it et al.} have estimated the magnetic field strength directly from the properties of the gas, such as vorticity and energy density. Dolag {\it et al.} started with a seed field at high redshift with a strength such that, at the present epoch, the field in clusters would be of the order of a few $\mu$G.  Miniati  assumed that the seed field was genererated through the Biermann battery mechanism, and is later rescaled to reproduce the measured magnetic field strength of galaxy clusters.  Donnert {\it et al.} have obtained the magnetic field in a way similar to Dolag {\it et al.}, but including additional effects at low redshifts, namely magnetic pollution. Despite the fact that Dolag {\it et al.} performed a constrained simulation, this may not be totally accurate due to the intrinsic properties of the simulation method. Because each of these cosmological simulations have their merits and problems, we will analyze the effects of all of them. It is worth mentioning that, beside the way the magnetic fields are obtained, these simulations also use different numerical techniques.

\begin{figure}[h!]
	\centering
	\includegraphics[width=0.83\columnwidth]{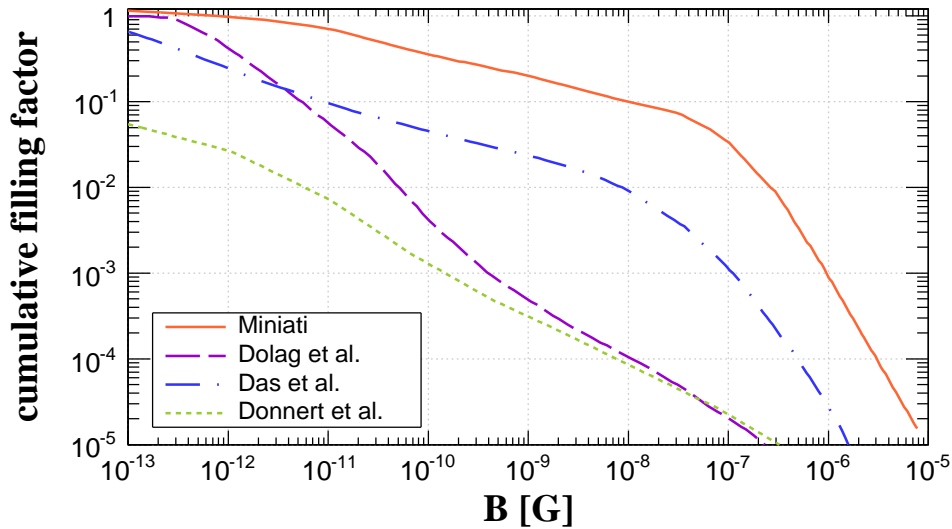}
	\caption{Cumulative filling factors for the cosmological simulations considered in this work. }
	\label{fig:fillingFactors}
\end{figure}

There are several cosmological simulations of the local universe beside the aforementioned ones. The method presented here can be applied to any cosmological simulation provided one has the filling factors distribution (or the magnetic field distribution), such as the ones shown in figure \ref{fig:fillingFactors}. We define the cumulative filling factors as the fraction of the volume that has a magnetic field strength higher than a given value. 

From a cosmological simulation one obtains a distribution of magnetic field strengths. The diffusion coefficient is calculated from the local field using equation \ref{eq:diffusion} assuming, in addition, a (constant) coherence length. By integrating equation \ref{eq:syro2} over the distribution of magnetic field strengths, we eliminate its $B$ dependence and obtain the value of $\lambda$ taking into account the inhomogeneity of the field.

\begin{figure}[h!]
	\centering
	\includegraphics[width=0.87\columnwidth]{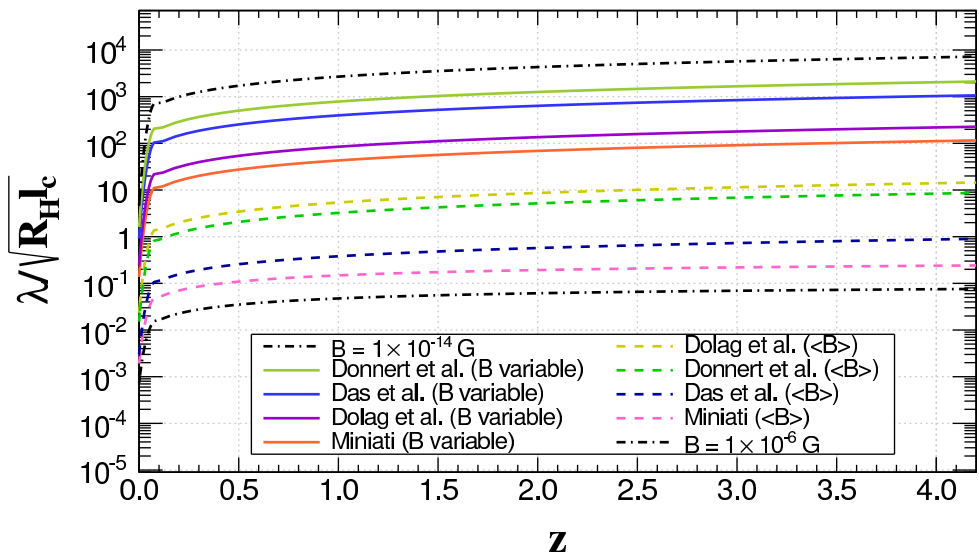}
	\caption{Volume averaged Syrovatskii variable for the magnetic fields from the large scale structure simulations (solid lines), in the case of constant magnetic fields equal to their average value (dashed lines), and two extreme cases (dotted dashed lines). The legend follows the same order as the curves, from top to bottom. This particular case is for $E/Z=10^{16}$ eV, $l_{c,0}=0.6$ Mpc, $m=1$, $\gamma=2.0$ and $z_{max}=4.0$.}
	\label{fig:syrovsz-lss}
\end{figure}

In figure \ref{fig:syrovsz-lss} we illustrate the redshift dependence of the volume averaged Syrovatskii variable for the aforementioned cosmological simulations, between $z$=0 and $z$=4, comparing them with the mean values of the corresponding magnetic field distributions, and with two extreme cases  ($B=1.0\times10^{-14}{\  }\mathrm{G}$ and $B=1.0\times10^{-6}{\  }\mathrm{G}$). These differences can also be seen if we use the root mean square value of the field for a given large scale structure. A summary of the mean and RMS value of each simulation are shown in table \ref{tab:distribprop}. It is important to stress that these values are not calculated from the actual MHD simulations, but from the filling factors distribution shown in figure \ref{fig:fillingFactors}. Also, for the sake of computational performance, we have restricted the magnetic fields to the range from $10^{-15}$ G to $10^{-5}$ G, considering it zero elsewhere.
\begin{table}[h!]
\centering
\caption{Mean and RMS values for the considered magnetic field distributions, in the range between 10$^{-15}$ G and 10$^{-5}$ G.\vspace{0.4cm}}
\begin{tabular}{c|cccc}
\hline
${\ }$ & Miniati & Dolag {\it et al.} & Das {\it et al.} & Donnert {\it et al.} \\
\hline
$\langle B \rangle$ [G] & $1.8\times10^{-8}$  & $5.5\times10^{-11}$ & $1.2\times10^{-9}$ & $6.3\times10^{-11}$\\
$B_{rms}$ [G] & $1.7\times10^{-7}$ & $1.5\times10^{-8}$ & $5.7\times10^{-8}$ & $1.7\times10^{-8}$
\end{tabular}
\label{tab:distribprop}
\end{table}

In figure \ref{fig:syrovsz-lss} we can clearly see that the assumption of a variable magnetic field implies a change in the value of the Syrovatskii variable and hence the spectrum. This is expected because only a small fraction of the volume is filled with magnetic fields with high values of $B$. Moreover the Syrovatskii variable when defined in units of length translates into the magnetic horizon. For instance, if we take $l_{c,0}$=1 Mpc and choose the two black (dotted dashed) lines from figure \ref{fig:syrovsz-lss}, we obtain a magnetic horizon of approximately 47 Gpc for $B$=10$^{-14}$ G and 60 Mpc for $B$=10$^{-6}$ G, taking into account the contribution of all sources up to $z_{max}=4$.

One should notice that all the calculations in this work depend on the assumption that the magnetic fields are turbulent, with diffusion coefficients given by equation \ref{eq:diffusion}, and with the magnetic field distributions from these particular cosmological simulations. Even though we consider the redshift evolution of the magnetic field ($B(z)=B_0(1+z)^{2-m}$), we do not follow the proper evolution of the whole cosmological simulation. Instead, we use the magnetic field strength at $z=0$ and extrapolate it to higher redshifts.

\section{Magnetic Suppression}

As mentioned before, if the term $F$ in equation \ref{eq:spectot1} is equal to 1, the spectrum does not depend on the modes of propagation and hence will be universal. We can define the suppression factor $G$ as the ratio between a given spectrum ($j(E)$) and the universal one ($j_0(E)$), i.e.
\begin{equation}
 G = \frac{j(E)}{j_0(E)}.
\label{eq:G2}
\end{equation}
Using equation \ref{eq:spectot1} we obtain the spectra for the cosmological simulations whose filling factors are shown in figure \ref{fig:fillingFactors}. Then we calculate the suppression factor G, shown in equation \ref{eq:G2}. The next step is to fit the suppression factor with the function:
\begin{equation}
 G(x) = \exp \left[ - \frac{(aX_s)^\alpha}{x^\alpha + bx^\beta}  \right],
 \label{eq:G}
 \end{equation} 
where $\alpha$, $\beta$, $a$ and $b$ are free parameters, $x$ is the ratio between the energy of the particle ($E$) and the average critical energy ($\left<E_{c,0}\right>$), and $X_s$ is given by
\begin{equation}
 X_s = \frac{d_s}{\sqrt{R_H l_c}},
  \label{eq:Xs}
\end{equation}
where $d_s = 3 / (4\pi n_s)$ is the average separation between the sources for a source density of $n_s$, and $R_H=c/H_0$ the  Hubble radius. We have assumed the source density to be constant over the evolution of the universe. It is important to mention that other functions may fit the suppression as well as, or even better than the one shown in equation \ref{eq:G}. Our choice was motivated by ref. \cite{mollerach2013}, and it proved itself to be adequate for our purposes. 

\begin{figure}[h!]
	\centering
	\includegraphics[width=0.48\columnwidth]{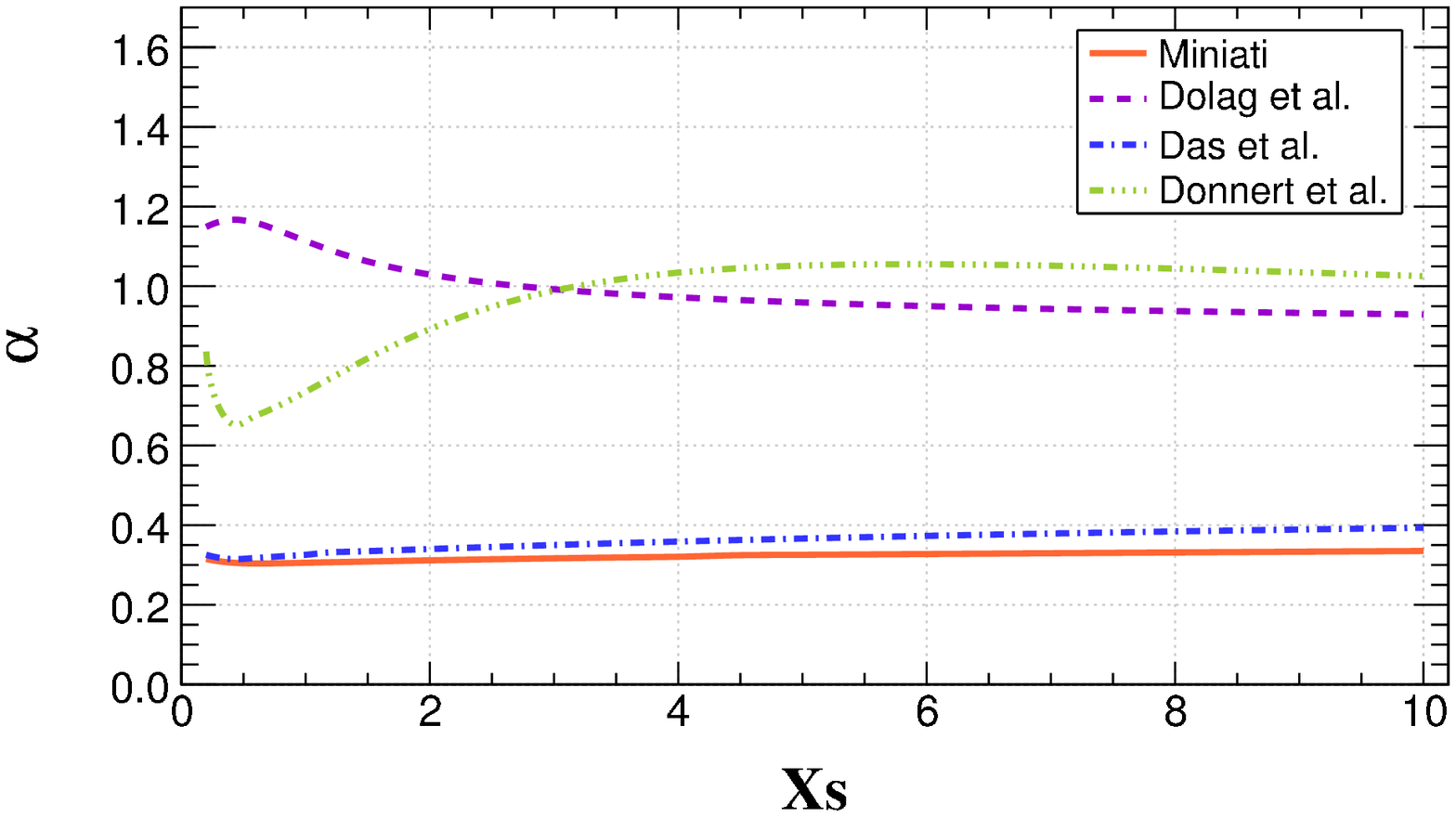}
	\includegraphics[width=0.48\columnwidth]{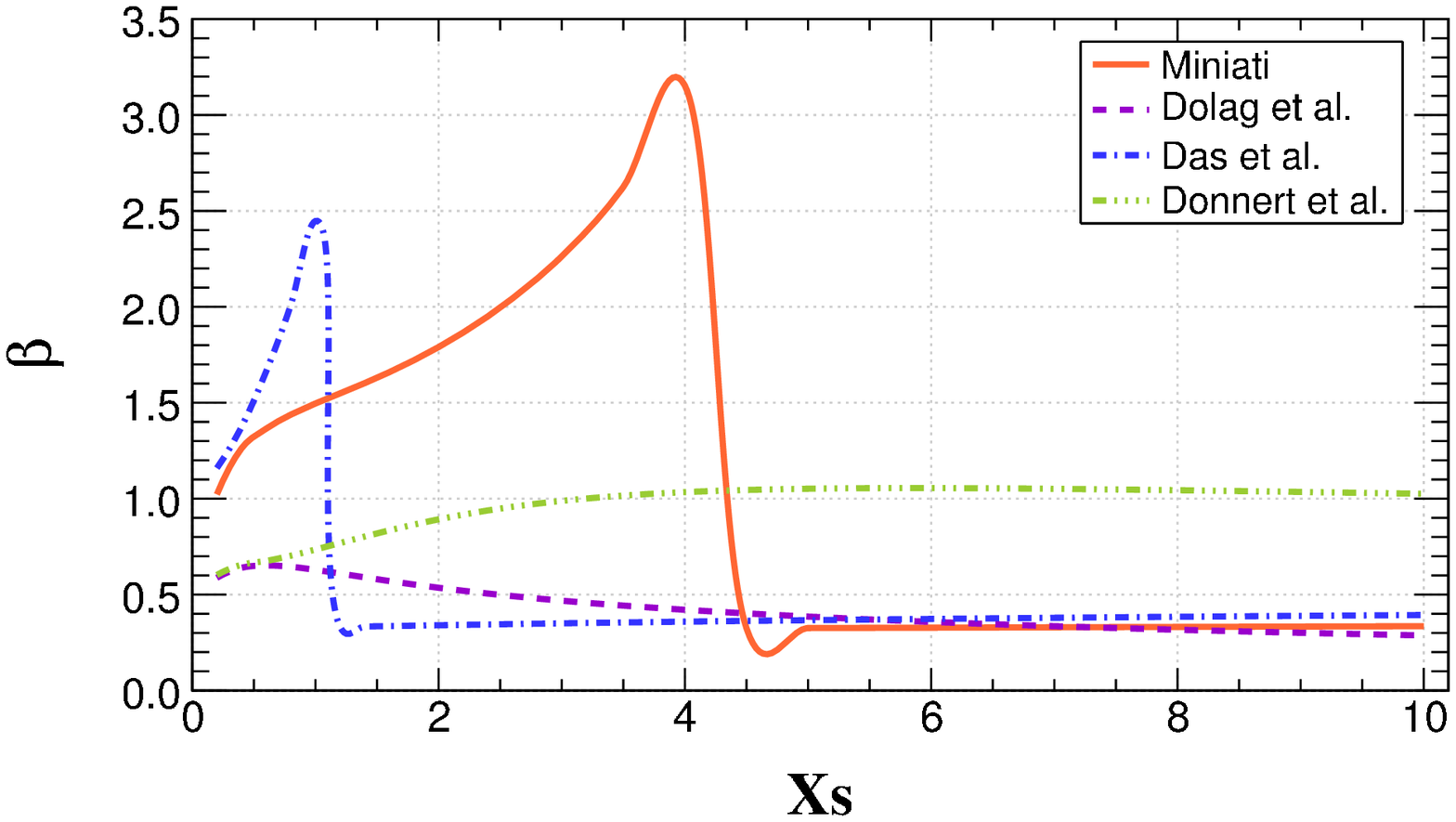}
	\includegraphics[width=0.48\columnwidth]{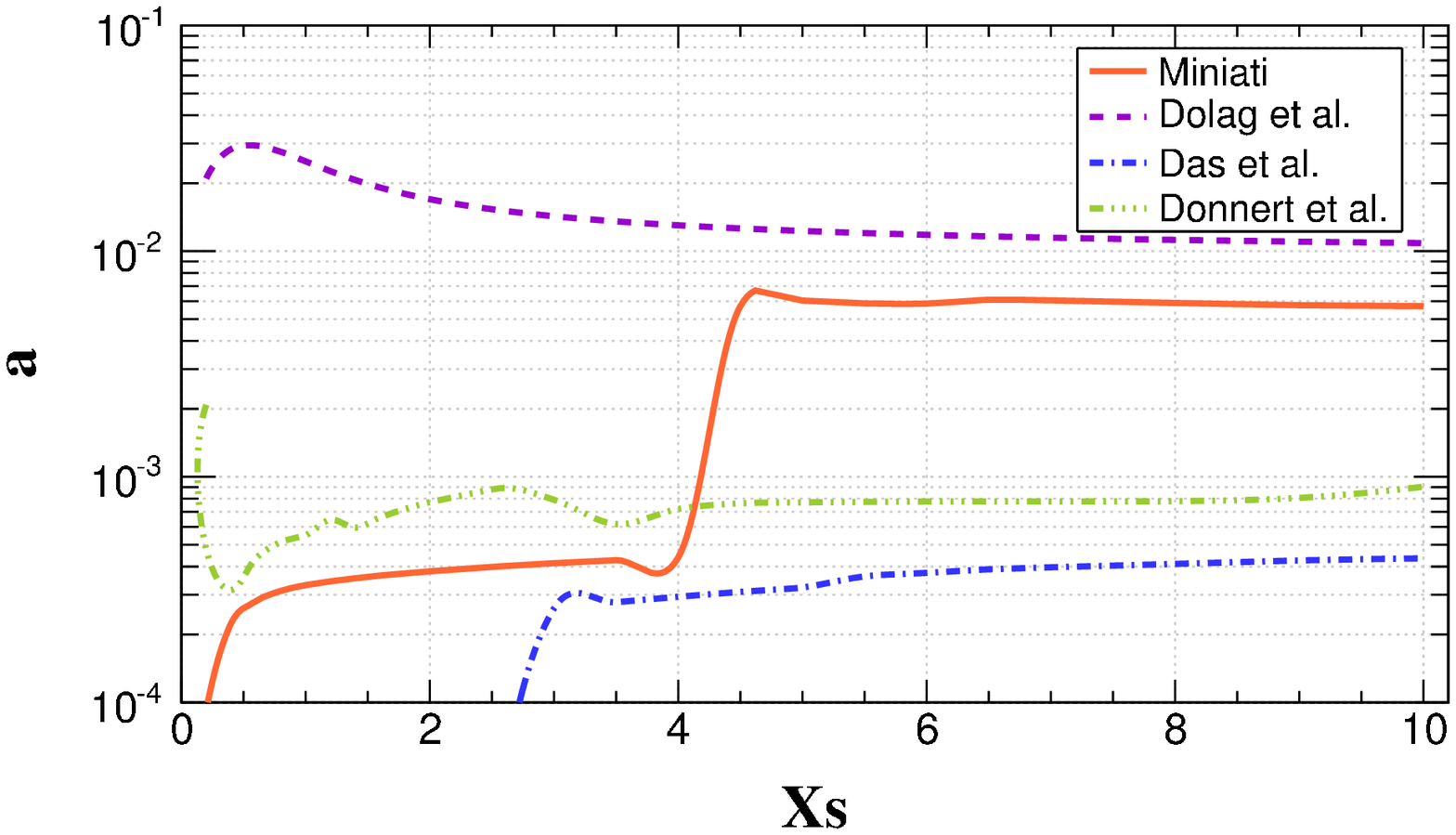}
	\includegraphics[width=0.48\columnwidth]{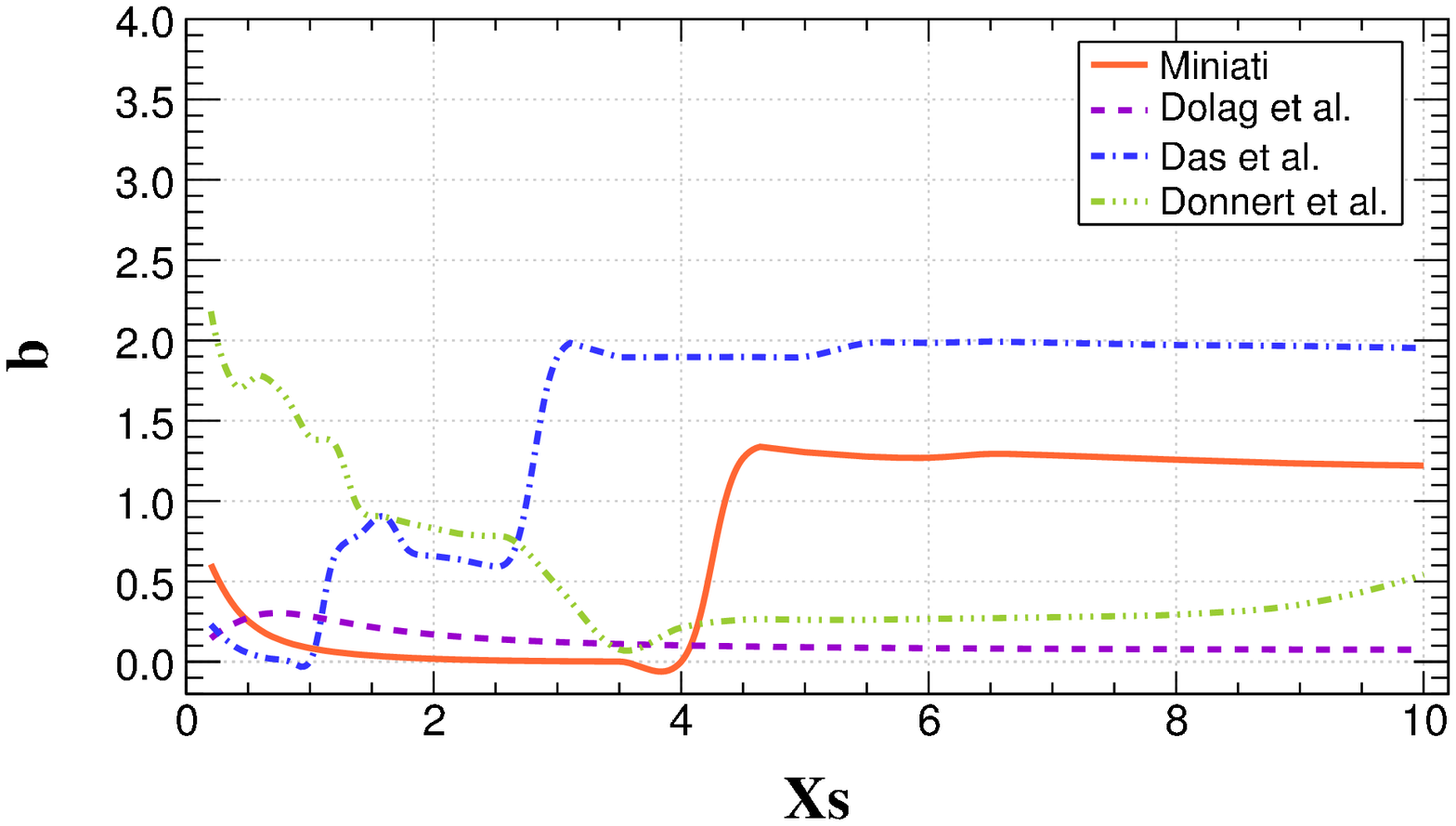}
	\caption{Values of the best fit parameters $\alpha$, $\beta$, $a$ and $b$ as a function of $X_s$ for $m=0$.}
	\label{fig:parameters}
\end{figure}

\begin{figure}[h!]
	\centering
	\includegraphics[width=0.48\columnwidth]{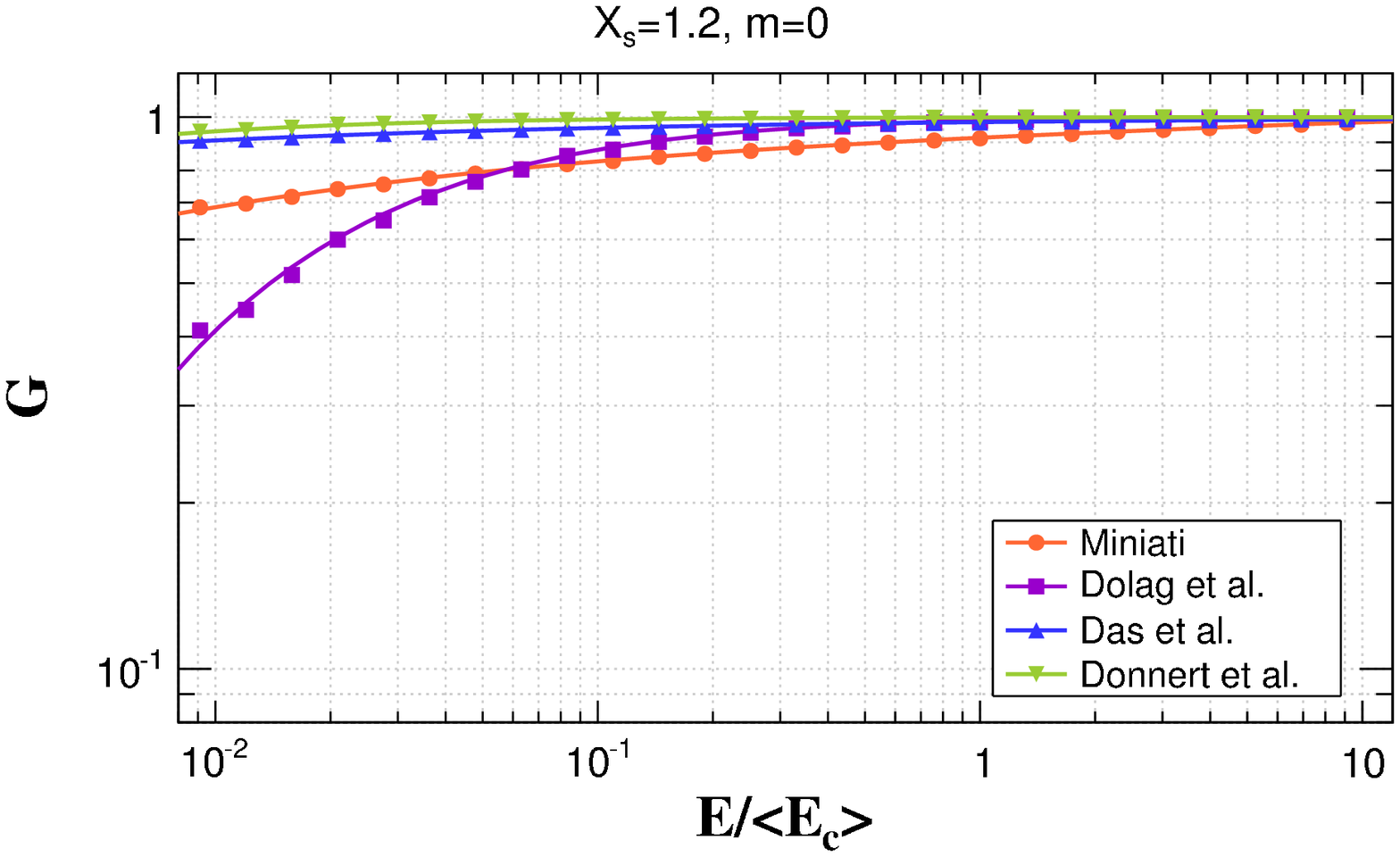}
	\includegraphics[width=0.48\columnwidth]{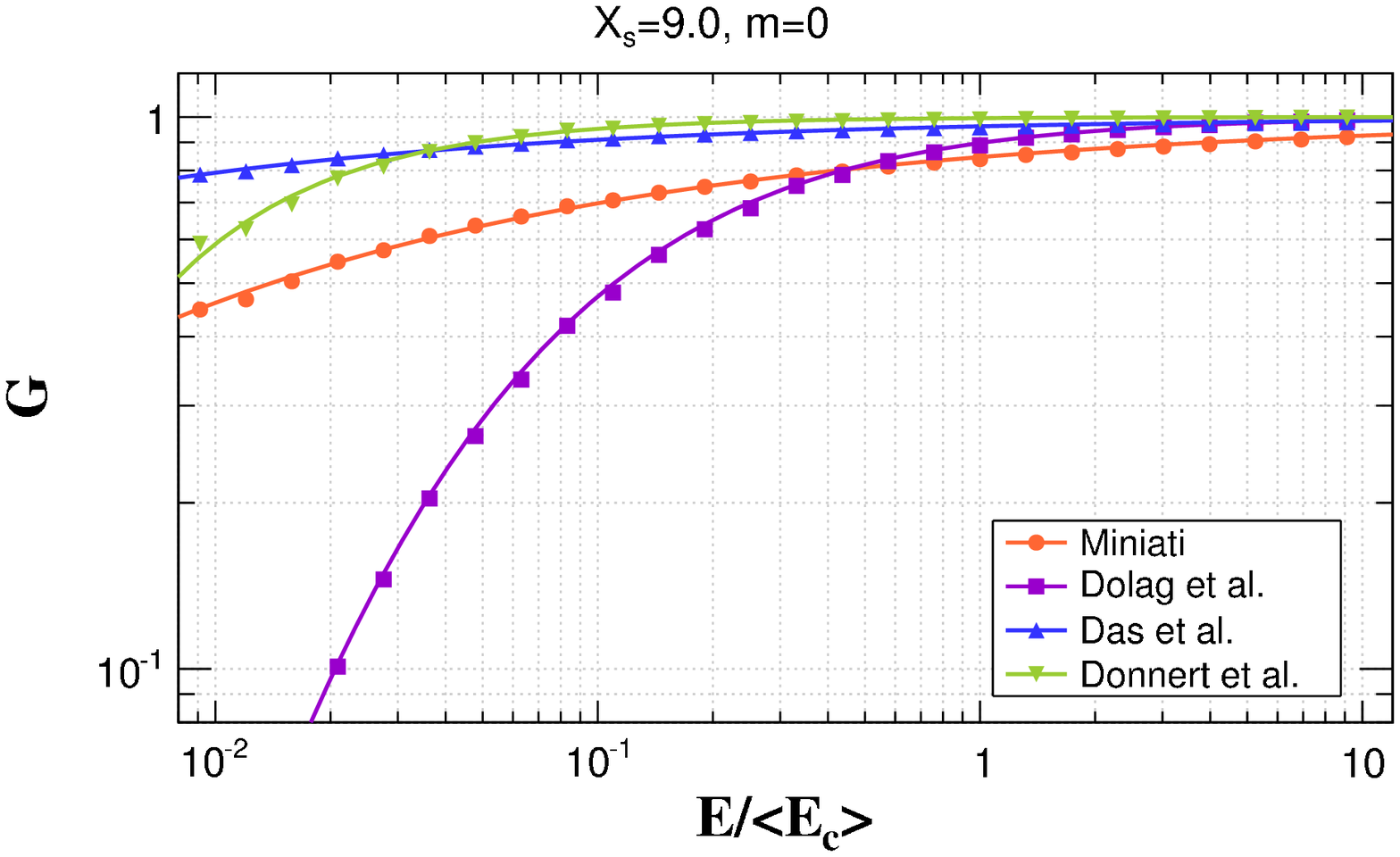}
	\caption{Suppression factor as a function of $x\equiv E/\langle E_{c,0}\rangle$. The markers are the suppression factors estimated using equation \ref{eq:G2}, whereas the lines correspond to the best fit values obtained using equation \ref{eq:G}, for $m$=0.}
	\label{fig:fits}
\end{figure}

The parameters of the fit vary for low values of $X_s$, but are practically constant for higher values. These results are summarized in figure \ref{fig:parameters}, and the fits can be seen in figure \ref{fig:fits}. They do not have a significant dependence on the spectral index of the source ($\gamma$) nor the cutoff energy ($E_{max}$), so they can be used generically. Also, since the suppression factor is written in terms of $E/\left<E_{c,0}\right>$, it will be the same for all nuclei with energy $E$ and rigidity $E/Z$. The parameter $m$, however, can affect the suppression factor, changing the values of the fit parameters, especially for the case of strong evolution ($m\gtrsim$2). A proper estimation of all the parameters for each values of $m$  would be required to obtain a more accurate description. Nevertheless, there are so many uncertainties involved (e.g. coherence length, power spectrum of the magnetic field, source density, source evolution), that the improvement of the fit parameters would not necessarily lead to better results. 

\begin{figure}[h!] 
	\centering
	\includegraphics[width=0.48\columnwidth]{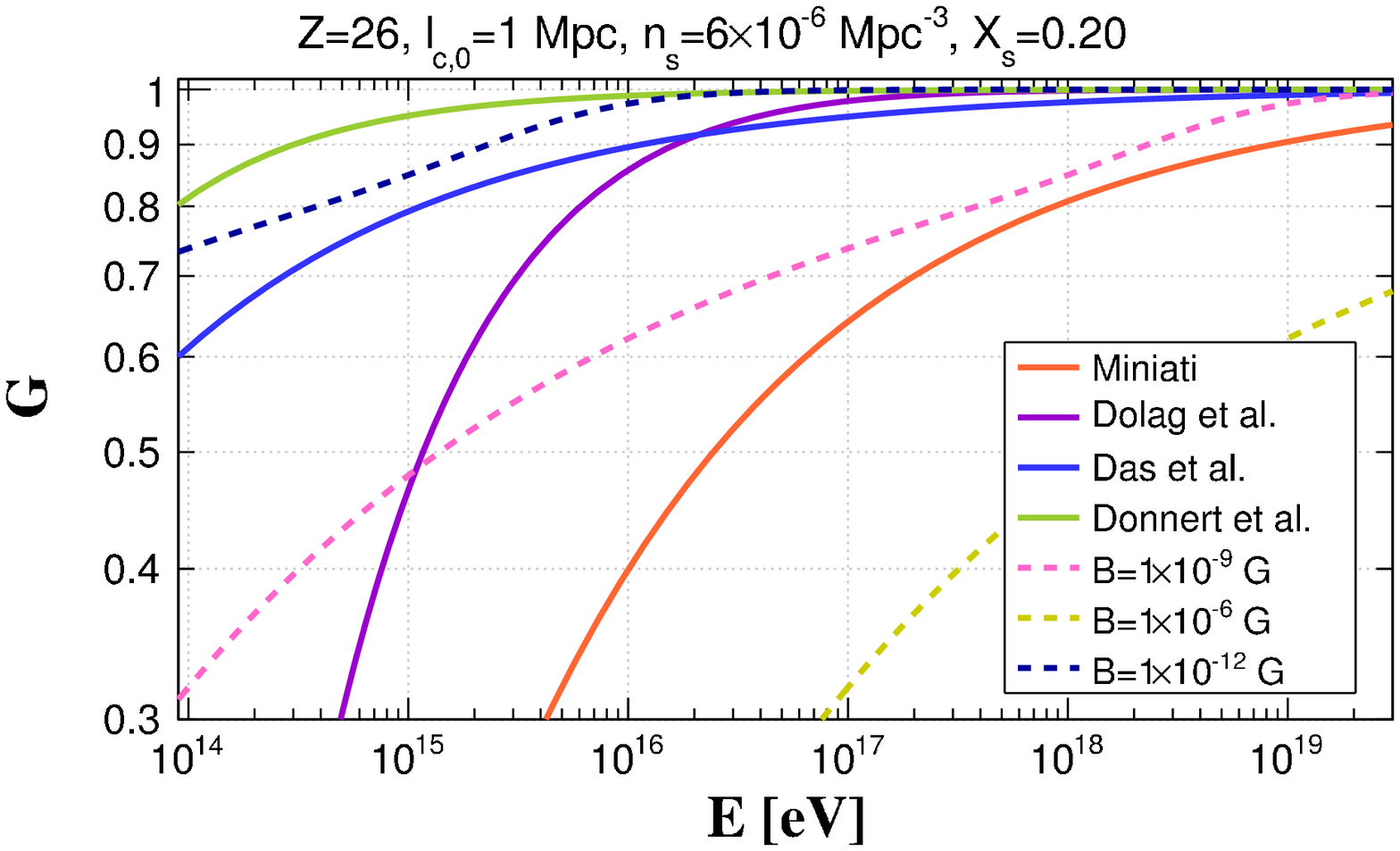}
	\includegraphics[width=0.48\columnwidth]{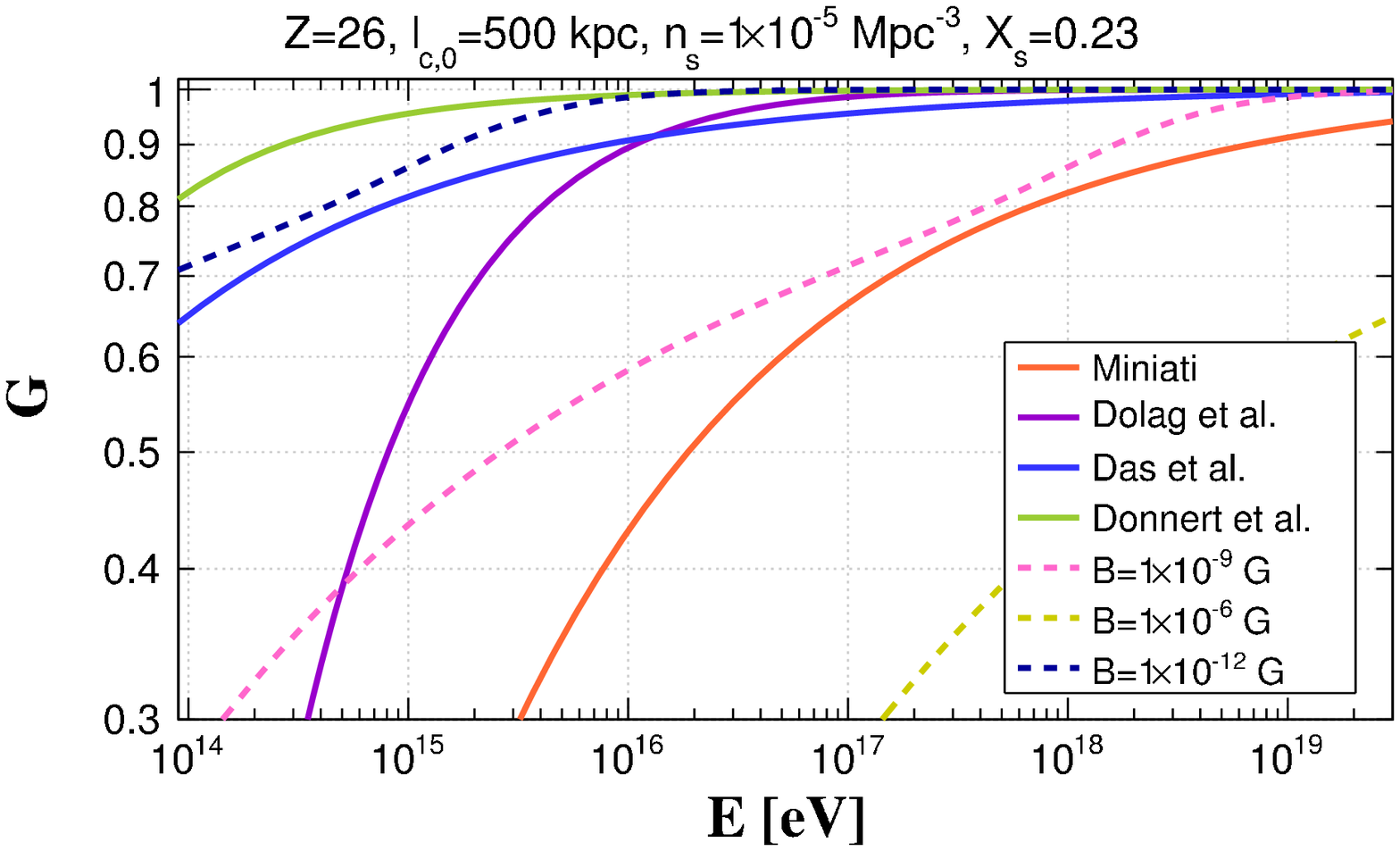}
	\includegraphics[width=0.48\columnwidth]{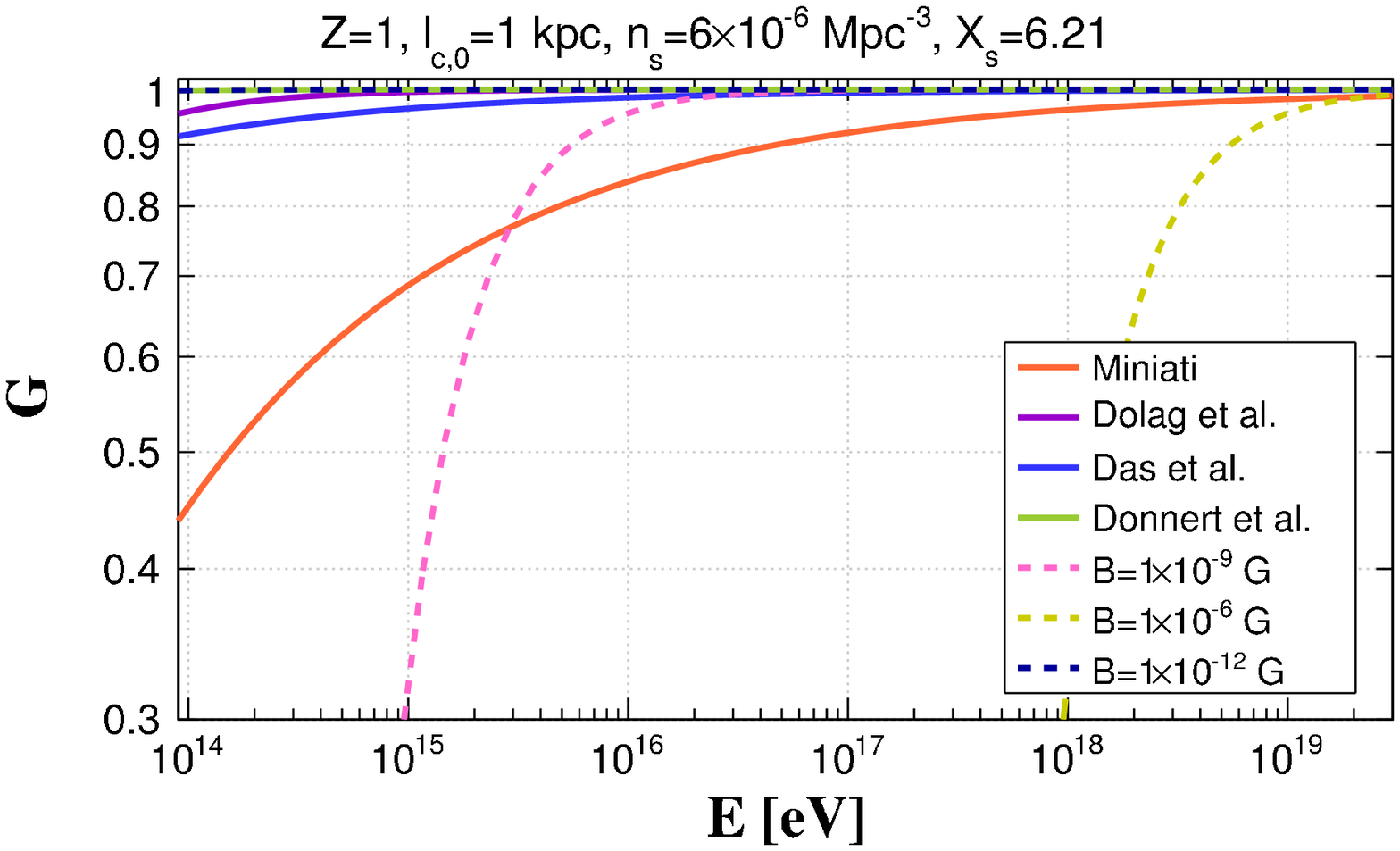}
	\includegraphics[width=0.48\columnwidth]{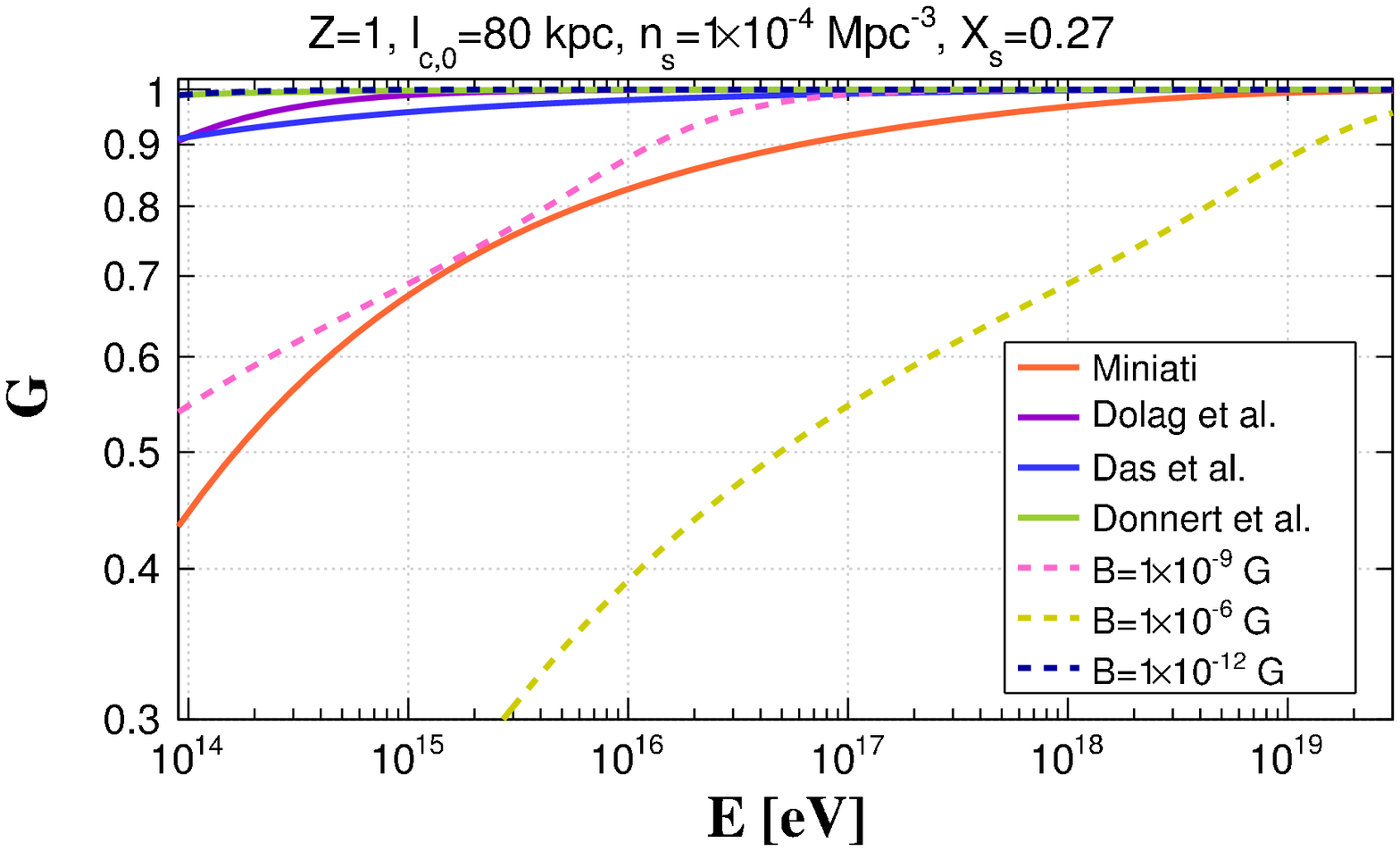}
	\caption{Suppression factor obtained from the parametrization as a function of energy. To generate this plot we have assumed the following parameters: $m=0$, $\gamma=2.0$, $E_{max}=100$ EeV and $z_{max}=4.0$. The panels in the top correspond to the case of $Z$=26, whereas the ones in the bottom to $Z$=1.}
	\label{fig:suppFactor}
\end{figure}

In figure \ref{fig:suppFactor} we show the suppression factors obtained from equation \ref{eq:G} as a function of the energy, for some combinations of $l_c$ and $n_s$, and hence $X_s$. For the sake of comparison, we present the results for the four magnetic field models previously described, together with three cases of constant magnetic field strengths, namely two extremes values ($B=$1 $\mu$G and  $B=$1 pG), and an intermediate one ($B=$1 nG). The suppression factors for the constant magnetic field cases were obtained assuming the parametrization presented in  ref. \cite{mollerach2013}. In this work the authors performed an analysis similar to the one here presented, but using a turbulent Kolmogorov magnetic field  with a fixed $B_{rms}$ of the order of a few nG. The spectrum in this case was calculated using equation \ref{eq:spectot1}, whereas we have used equation \ref{eq:spectot2}, since our assumption is a distribution of magnetic field strengths. Hence for the case of a turbulent magnetic field with fixed $B_{rms}$ we have adopted the best fit parameters from ref. \cite{mollerach2013} : $\alpha$=1.43, $\beta$=0.19, $a$=0.20, $b$=0.09. 

It is possible to notice a few interesting features in figure \ref{fig:suppFactor}. The first one is that, beside the constant magnetic field cases, the only model in which we obtain a suppression of the order of 20\% around 1 EeV is in the Miniati case, assuming a heavy composition ($Z$=26), as shown in the top panels. The top right panel is almost identical to the top left one, with approximately the same $X_s$, but in this case the source density is higher and coherence length lower. In the two bottom panels, which correspond to a purely protonic composition, it is possible to see that if the mass composition of the cosmic rays is light then the low energy suppression due to diffusion would take place below $E\lesssim$10$^{16}$ eV.

We can estimate the energy $E_e$ at which we have a suppression of $e^{-1} \approx 0.37$. For that we start with equation \ref{eq:G} and calculate $G(E_e/\langle E_{c,0} \rangle)=1/e$, obtaining
\begin{equation}
  E_e^\alpha + b E_e^\beta \langle E_{c,0} \rangle^{\alpha - \beta} = (a X_s \langle E_{c,0} \rangle)^\alpha.
  \label{eq:Ee}
\end{equation}
Similarly we can find the coherence length the corresponding coherence length for $E_e$:
\begin{equation}
(a d_s cZe\langle B_0 \rangle)^\alpha R_H^{-\alpha/2} l_{c,0}^{\alpha/2} - b (cZe\langle B_0 \rangle)^{\alpha-\beta} E_e^\beta = E_e^\alpha . 
\label{eq:limlc}
\end{equation}

\begin{figure}[h!]
	\centering
	\includegraphics[width=0.83\columnwidth]{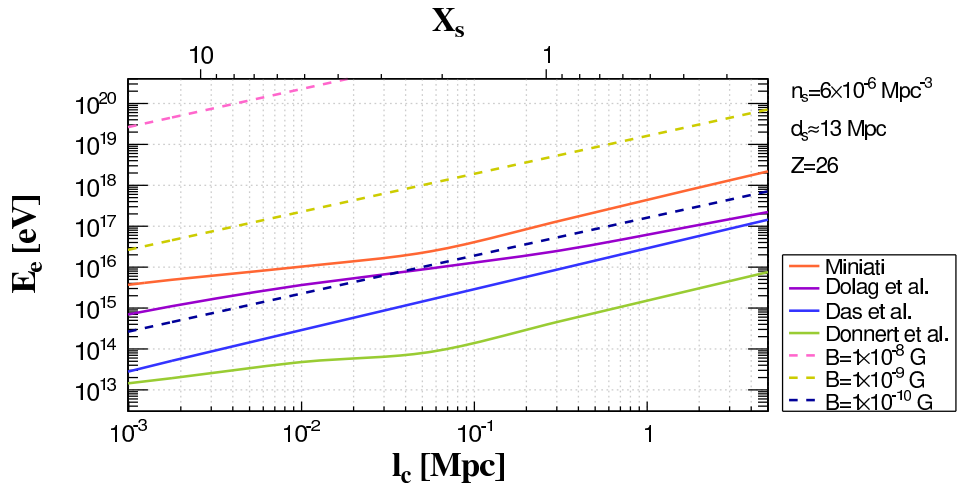}
	\caption{Energy for which the flux is suppressed to $1/e$ of its former value, as a function of the coherence length. Dashed lines correspond to the cases of constant magnetic field strength and solid lines to magnetic fields from the cosmological simulations indicated in the legend. This particular case is for $n_s$=6$\times$10$^{-6}$ Mpc$^{-3}$ and $Z$=26.}
	\label{fig:Eelc}
\end{figure}

Using equation \ref{eq:Ee} we can estimate the energy for which the flux is suppressed to $1/e$ of its original value due to diffusion of particles in extragalactic magnetic fields, assuming that the sources are uniformly distributed. This is shown in figure \ref{fig:Eelc} for a source density of 6$\times$10$^{-6}$ Mpc$^{-3}$ for the case of iron. Since this source density is a lower limit \cite{auger2013a}, and for iron the suppression is stronger, figure \ref{fig:Eelc} can be understood as an upper limit for a suppression of the flux to $1/e$ of its former value, due to diffusion. One should bear in mind that if nuclear photodisintegration occurs, and it very likely will,  the curves displayed in this figure will be shifted to even lower energies.

In figure \ref{fig:Eelc} we notice that the energy $E_e$ increases with the coherence length $l_{c}$ in the range considered. To constrain the constant magnetic field scenarios displayed in this figure we have again used the parameters $\alpha$, $\beta$, $a$ and $b$ from ref. \cite{mollerach2013}, which are the same for all values of $X_s$. In our case, for the inhomogeneous magnetic field models, we assumed that these parameters vary with $X_s$ to obtain better fits of the analytical model.

The low energy suppression is stronger for $l_c \sim $ Mpc. This dependence can be understood by analyzing the behavior of the diffusion coefficient, shown in equation \ref{eq:diffusion}, for different values of $l_{c}$. The critical energy $E_c$ is proportional to $l_{c}$, as displayed in equation \ref{eq:Ec}. Therefore, at a given energy, for small values of $l_{c}$, the term proportional to $E^2$ in equation \ref{eq:diffusion} would dominate, implying $D \propto l_c^{-1}$, and thus $\lambda \propto l_c^{-1/2}$.  This leads to a flux suppression and a critical energy both of which increase with $l_c$. Similarly, if $l_c$ is large, $D \propto l_c^{2/3}$ and hence $\lambda \propto l_c^{1/3}$, implying that $E_e$ should decrease with $l_c$. However, this last effect is not visible in figure \ref{fig:Eelc}, due to the limited range of parameters covered in our analysis, tuned to encompass typical values of $l_c$ found in the literature, and allow an analytical description of the suppression through equation \ref{eq:G}.

\section{Discussion}

Our results indicate a very weak suppression of the flux of cosmic rays at 10$^{18}$ eV, which starts to become more pronounced at $E\lesssim$ 10$^{16}$ eV, depending on the magnetic field model, its coherence length, and the charge of the particle. This result is in qualitative agreement with ref. \cite{kotera2008}, in which three other MHD simulations were used and an analysis similar to ours was performed. In this work the authors observed the presence of a magnetic horizon for energies below $\sim$ 10$^{17}$ eV, which roughly corresponds to our upper limit on the energy the suppression sets in, for the most optimistic choice of parameters. Even though we did not analyze the MHD simulations used in this reference, their filling factors distributions are lower compared to the Miniati case, particularly for $B\gtrsim$100 nG. Since a larger contribution from higher magnetic fields would cause a stronger suppression, we can say that the Miniati scenario can be taken as a limiting case for the ones presented in ref. \cite{kotera2008}.

We have adopted a uniform source distribution, but other possibilities could be easily accounted for by changing equation \ref{eq:ri}, since the spectrum is the superposition of the fluxes of individual sources. At energies below 1 EeV, the spectrum is dominated by  the contribution of distant sources. At cosmological distances the distribution of sources is probably close to uniform. Therefore, unless the sources are clustered and their distribution is highly non uniform, the results here obtained would not change significantly.
 

To have diffusion from the nearest source the diffusion length ($l_D = 3D / c$) should be smaller than the distance of this source, i. e., $l_D < d_s$. This effect may be dominant depending on the source distance and luminosity. Therefore, a more precise calculation of the suppression would need to take into account the inhomogeneity of the magnetic field in scales comparable to the distance of the nearest source. As long as the filling factors distribution for a volume containing both the nearest source and Earth has the same shape as the ones we considered, the results here presented will hold. This is not true in scales comparable to the size of the structures. Moreover it may not be valid if both the observer and the source lie within the same filament, due to the higher magnetic field strengths in these regions compared to the voids. This would change the magnetic field distribution, shifting its mean value toward higher values of $B$, possibly spawning a stronger suppression in the observed flux if the luminosity of the source is high enough. 

Another assumption made in this work is that the coherence length if fixed, which is hardly realistic. The analysis that we have presented can be easily extended to incorporate coherence length distributions. However, the actual distribution of coherence lengths is not well known and can vary in different regions of the universe. In this work we analyzed coherence lengths between $\sim$ 10$^{-3}$ and 10 Mpc, which are typical values in the literature \cite{carilli2002,ryu2011,neronov2013}. Regardless of the shape of this distribution, as long as the contribution of coherence lengths outside the studied range is negligible, $E_e$ for $l_c$=10 Mpc  can still be taken as an upper limit for the magnetic suppression.

Sigl \cite{sigl2007} has shown that the confinement of particles around the source position may play a role in the low energy suppression. If this is true, there might be an additional contribution, due to the fact that the magnetic field would be much higher around the source than elsewhere. This would occur if the following condition is satisfied
\begin{equation}
 B \gtrsim \sqrt{\frac{l_c}{10{\  }\mathrm{kpc}}} \left( \frac{L}{100{\  }\mathrm{  kpc}} \right){\  } \mu\mathrm{G},
\end{equation}
where $L$ represents the characteristic scale of the magnetized region. If the magnetic fields have stochastic nature with a relatively low RMS value ($B \lesssim$ 100 nG), this confinement would not be expected. 

The neglected energy loss processes are not so relevant at $E\lesssim$ EeV, as mentioned earlier. At energies of a few EeV it has been shown in ref. \cite{mollerach2013} that the effect of pair production would be small, slightly shifting the low energy suppression to higher energies. This effect is also negligible for $E \lesssim$ EeV. 

The results here obtained are for the range of 10$^{-3} <$ x$\equiv E/\langle E_c \rangle$ $<$ 10. Although not systematically, we have tested the fit using equation \ref{eq:G} in the range of $10^{-4} < x < 100$. For $x\sim100$ the fit describes the suppression well. For $x\sim$10$^{-4}$, however, equation \ref{eq:G} no longer fits satisfactorily the values calculated using equation \ref{eq:G2}.

In this work we have used the diffusion approximation to describe the effects of extragalactic magnetic fields on the cosmic ray spectrum and composition. This approximation is not valid for $D\gtrsim cR_H$ because the diffusive propagation speed over a time scale $t_H\equiv R_H c^{-1}$ is $\simeq(D/t_H)^{1/2}$ which would then exceed the speed of light\footnote{A phenomenological approach to the problem of superluminal diffusion of cosmic rays can be found in ref. \cite{aloisio2009}.} and hence be unphysical. With $D$ given by eq. \ref{eq:diffusion} this typically happens for $B\lesssim0.01$ nG. In summary, if $D\gtrsim R_Hc$ and/or $D/\ell_D\gtrsim c$ the diffusion approximation can no longer be applied and one would have to adopt a full numerical Monte Carlo simulation of trajectories. 

In ref. \cite{mollerach2013} it was argued that the low energy suppression could be relevant for the propagation of cosmic rays at $\sim$ EeV energies, which is in contrast with our results. This discrepancy is due to the oversimplified assumption of a turbulent magnetic field with constant strength. In the case of inhomogeneous magnetic fields the contribution of the voids is dominant, lowering the average field strength. 

Many authors \cite{hooper2010,taylor2011,allard2012,aloisio2013b} have recently attempted to obtain models that can simultaneously describe the measured spectrum and mass composition of UHECRs.
In ref. \cite{aloisio2013b} it was shown that it is only possible to perform combined spectrum-composition fits of the Auger data for $E\gtrsim$5$\times$10$^{18}$ eV, due to the hard spectral indexes required ($\gamma \sim$1.0-1.6). Hard spectra are incompatible with the standard acceleration paradigm, in which particles are accelerated in non-relativistic shocks through a first order Fermi mechanism, process which leads to $\gamma \sim$ 2. This is also incompatible with relativistic shock acceleration, which predicts $\gamma \sim$2.3. Hard spectral indexes ($\gamma\lesssim 2$) are predicted in many other acceleration models such as the ones in which UHECRs are accelerated by  magnetars \cite{arons2003} or young pulsars \cite{fang2012,fang2013}.

We have shown that for the magnetic field models studied, the low energy suppression of the extragalactic flux is mild at EeV energies, becoming more relevant at energies $\lesssim$10$^{17}$ eV. Since the spectral index of the source and the existence of a magnetic horizon are connected, understanding the low energy suppression is important for identifying the sources of UHECRs. In terms of cosmic ray observables, harder spectral indexes could lead to an overproduction of secondary protons for  $E\sim$10$^{18}$ eV. This same effect could be mimicked by considering softer injection spectra which are effectively hardened during propagation by the effect of the magnetic suppression, as shown in ref. \cite{mollerach2013}. In the context of our work, if the scale of inhomogeneity of the cosmic web is of the order of the distance of the nearest sources, i. e., if there are no dominant nearby sources, the suppression is very low at EeV energies, as can be seen in figure \ref{fig:Eelc}. In the Miniati model, which has the higher mean magnetic field, the suppression would become significant only at $E\lesssim$10$^{17}$ eV for the most optimistic choice of parameters. The main implication of this is that the combined spectrum-composition fits would again favor scenarios in which the sources have hard injection spectrum.

\section{Conclusions}

We have derived an approximation for the magnetic suppression of the cosmic ray flux from distant sources for $E\lesssim$ Z EeV. This suppression will occur when the propagation time of particles in cosmic magnetic fields are comparable to the age of the universe. Our result extends the previous work from Mollerach \& Roulet \cite{mollerach2013} by considering a magnetic field distribution, rather than a constant value. We have assumed a Kolmogorov magnetic field with strengths distributed according to cosmological simulations of the local universe done by Miniati, Dolag {\it et al.}, Das {\it et al.} and Donnert {\it et al.}. Since in these simulations most of the volume of is filled by voids, then the low magnetic field strengths from these regions will be preponderant to the propagation of cosmic rays, dominating over the high values corresponding to clusters of galaxies and filaments. This assumption will imply a milder suppression compared to the case of a constant magnetic field or, depending on the values of the coherence length, none.  

The approximation here presented is volume averaged and does not reflect local effects such as the nearby distribution of magnetic fields. For instance, if both the source and the observer lie within the same filament, or if the source is in a highly magnetized region, this approximation may no longer be valid, depending on the distance and luminosity of the nearest source.  To account for these effects three dimensional simulations with a full Monte Carlo approach are needed. Nevertheless, considering only the extragalactic component, at energies $\lesssim 10^{18}$ eV the bulk of the flux is composed by particles from distant sources.

The method to estimate the suppression can be easily adapted to any other cosmological simulation provided that one has its magnetic field distribution. However it is important to bear in mind that the parametrizations for the suppressions for the different magnetic field models here considered are very rough and other effects such as the structure of the magnetic fields may be relevant. Moreover, many parameters such as the source density and the coherence length are set by hand. The actual value of the coherence length of extragalactic magnetic fields is not well established. It is intrinsically connected to the cosmological magnetogenesis, and can be a distribution rather than a constant value.

An improvement in the method here presented would be to use the power spectrum of the cosmological simulation to obtain the diffusion coefficient, instead of using the approximation of a Kolmogorov field. Also, we have extrapolated the magnetic field distribution at present time up to higher redshifts, which is a very crude approximation, given that the overall evolution of the simulation volume is not as simple as $B_0(1+z)^{2-m}$ when structure formation and MHD effects other than  adiabatic compression are taken into account. Another improvement would be to consider a distribution of coherence lengths, possibly, but not necessarily coupled to the magnetic field strength. 

We have also derived model dependent upper limits for the suppression of the flux due to magnetic horizon effects. These results show that in the absence of nearby dominant sources the extragalactic component can be significantly suppressed only below $E\lesssim$ 10$^{17}$ eV, provided that the coherence length of the extragalactic magnetic fields is smaller than a few Mpc.

\acknowledgments

This work was supported by the Deutsche Forschungsgemeinschaft (DFG) through the Collaborative Research Centre SFB 676 ``Particles, Strings and the Early Universe'' and by BMBF under grants  05A11GU1 and 05A14GU1. RAB acknowledges the support from the Forschungs- und Wissenschaftsstiftung Hamburg through the program ``Astroparticle Physics with Multiple Messengers''.  Furthermore, we acknowledge support from the Helmholtz Alliance for Astroparticle Phyics (HAP) funded by the Initiative and Networking Fund of the Helmholtz Association. We also thank Silvia Mollerach for valuable comments.

\appendix

\section{Diffusion equation in an expanding universe}
\label{app:appendixA}

In this appendix we derive the diffusion equation as done by Berezinsky and Gazizov \cite{berezinsky2006a}, with some adjustments to suit our purposes.

Let $n(\vec{x},t)$ be the particle density. This quantity can be written in another basis as $n(\vec{r},t)$, where $x(t) = a(t) r$, with $a$ being the scale factor of the universe.
 Consider an expanding sphere with radius $r$ at time $t$. The diffusive flux can be written as
 \begin{equation}
 	\vec{j} = -D(\vec{x},t) \vec{\nabla}_x n(\vec{x},t),
	\label{eq:fluxj}
 \end{equation}
 where the subscript $x$ indicates that the operator $\vec{\nabla}$ is written in the basis $x$. 

To derive the diffusion equation we start by writing the continuity equation in its integral form:
\begin{equation}
	\frac{d}{dt} \int \limits_{V(t)} dV n(\vec{x},t) = - \oint \limits_{S(t)} \vec{j} \cdot dS,
\end{equation}
where $S(t)$ is the expanding sphere corresponding to the universe.
Replacing the value of $\vec{j}$ by equation \ref{eq:fluxj} and applying Gauss's divergence theorem we obtain:
\begin{equation}
	\frac{d}{dt} \int \limits_{V(t)} dV n(\vec{x},t) = \int \limits_{V(t)} dV \vec{\nabla}_x \cdot \left[  D(\vec{x},t) \vec{\nabla}_x  n(\vec{x},t) \right].
	\label{eq:continuity}
\end{equation} 
Now we calculate the left-hand side of this equation, assuming that at time $t$ the universe has a volume $V(t)$ and at time $t+\delta t$ volume $V+\delta V$:
\begin{equation}
	\frac{d}{dt} \int \limits_{V(t)} dV n(\vec{x},t) = \int \limits_{V(t)} dV\frac{ \partial n(\vec{x},t)}{\partial t} + n(\vec{x},t)  \frac{d}{dt} \left( \delta V\right).
\end{equation}
$V(t)$ is written in the $(\vec{x},t)$ basis. Writing the second term in the right-hand side of this equation as a function of the comoving volume ($V(t) = a^3(t) \mathcal{V}$) we get:
\begin{equation}
n(\vec{x},t) \frac{d}{dt} \left( \delta V \right) = n(\vec{x},t) \frac{d}{dt}\left( a^3 \delta \mathcal{V} \right) = 3 n(\vec{x},t)  H(t) a^3 \delta \mathcal{V} = 3H(t) n(\vec{x},t)   \delta V.
\label{eq:dVexpansion}
\end{equation}
Combining equations \ref{eq:continuity} and \ref{eq:dVexpansion} we have
\begin{equation}
	\int \limits_{V(t)} dV \frac{\partial n(\vec{x},t)}{\partial t} + 3H(t) n(\vec{x},t) \delta V =
	\int \limits_{V(t)} dV \vec{\nabla}_x \cdot \left[  D(\vec{x},t) \vec{\nabla}_x  n(\vec{x},t) \right].
\end{equation}
Changing the basis of the term inside the integral in the right-hand side of the equation to $(\vec{r},t)$, one obtains:
\begin{dmath}
	\vec{\nabla}_x \cdot \left[  D(\vec{x},t) \vec{\nabla}_x  n(\vec{r},t) \right] = \vec{\nabla}_x D(\vec{x},t) \cdot \vec{\nabla}_x n(\vec{x},t) + D(\vec{x},t) \nabla^2_x n(\vec{x},t) =  \frac{1}{a^2(t)}  \vec{\nabla}_r D(\vec{r},t) \cdot \vec{\nabla}_r n(\vec{r},t) + \frac{1}{a^2(t)} D(\vec{r},t) \nabla^2_r n(\vec{r},t)
\end{dmath}
Because $\delta V$ is small we can write:
\begin{equation}
  \frac{\partial n(\vec{r},t)}{\partial t} + 3H(t) n(\vec{r},t) = \vec{\nabla}_r D(\vec{r},t) \cdot \vec{\nabla}_r n(\vec{r},t) + \frac{1}{a^2(t)} D(\vec{r},t) \nabla^2_r n(\vec{r},t).
\end{equation}
To this equation we add a source term $Q(E,t)$, which is the density of particles with energy $E$ at a given time $t$:
\begin{equation}
	\mathrm{source{\  }term} = \frac{Q(E,t)}{a^3(t)} \delta^3(\vec{r}-\vec{r_g}).
\end{equation}
We also have to account for energy losses, so we add an additional term:
\begin{equation}
		\mathrm{energy{\  }loss{\  }term} =  \frac{\partial}{\partial E}  \left[ n(E,\vec{r},t) b(E,t) \right] .
\end{equation}
We can finally write the diffusion equation in an expanding universe taking into account energy losses:
\begin{dmath}
	 \frac{\partial}{\partial t} n(E,\vec{r},t) - \frac{\partial}{\partial E}  \left[ n(E,\vec{r},t) b(E,t) \right] + 3H(t) n(E,\vec{r},t) - \vec{\nabla} D(E,\vec{r},t) \cdot \vec{\nabla} n(E,\vec{r},t) - \frac{D(E,\vec{r},t)}{a^2(t)} \nabla^2 n(E,\vec{r},t) = \frac{Q (E,t)}{a^3(t)} \delta^3(\vec{r}-{\vec{r_g}}).
	 \label{eq:diffEqComplete}
\end{dmath}

Notice that if  $D(\vec{r},t) \nabla^2 n(\vec{r},t) \gg  \vec{\nabla}D(\vec{r},t) . \vec{\nabla}n(\vec{r},t)$ the diffusion coefficient can be considered position independent, hence equation \ref{eq:diffEqComplete} reduces to the one obtained in ref. \cite{berezinsky2006a}, whose solution is given by equation  \ref{eq:specss1}.

\bibliography{references}

\providecommand{\href}[2]{#2}\begingroup\raggedright\begin{thebibliography}{10}

\bibitem{auger2013b}
{Pierre Auger Collaboration}, {\it {Constraints on the Origin of Cosmic Rays
  above 10$^{18}$ eV from Large-scale Anisotropy Searches in Data of the Pierre
  Auger Observatory}},  {\em The Astrophysical Journal. Letters} {\bf 762}
  (Jan., 2013) L13, [\href{http://xxx.lanl.gov/abs/1212.3083}{{\tt
  arXiv:1212.3083}}].

\bibitem{auger2010}
{Pierre Auger Collaboration}, {\it {Measurement of the depth of maximum of
  extensive air showers above 10$^{18}$ eV.}},  {\em Physical Review Letters}
  {\bf 104} (Mar., 2010) 091101, [\href{http://xxx.lanl.gov/abs/1002.0699}{{\tt
  arXiv:1002.0699}}].

\bibitem{hires2010}
{High Resolution Fly's Eye Collaboration}, {\it {Indications of
  Proton-Dominated Cosmic-Ray Composition above 1.6 EeV}},  {\em Physical
  Review Letters} {\bf 104} (Apr., 2010) 161101,
  [\href{http://xxx.lanl.gov/abs/0910.4184}{{\tt arXiv:0910.4184}}].

\bibitem{tinyakov2014}
P.~Tinyakov, {\it {Latest results from the Telescope Array }},  {\em Nuclear
  Instruments and Methods in Physics Research Section A: Accelerators,
  Spectrometers, Detectors and Associated Equipment} {\bf 742} (2014), no.~0 29
  -- 34. 4th Roma International Conference on Astroparticle Physics.

\bibitem{kascade2011}
{KASCADE-Grande Collaboration}, {\it {Kneelike Structure in the Spectrum of the
  Heavy Component of Cosmic Rays Observed with KASCADE-Grande}},  {\em Physical
  Review Letters} {\bf 107} (Oct., 2011) 171104,
  [\href{http://xxx.lanl.gov/abs/1107.5885}{{\tt arXiv:1107.5885}}].

\bibitem{peters1961}
B.~Peters, {\it Primary cosmic radiation and extensive air showers},  {\em Il
  Nuovo Cimento} {\bf 22} (1961), no.~4 800--819.

\bibitem{deligny2014}
O.~Deligny, {\it {Cosmic rays around : Implications of contemporary
  measurements on the origin of the ankle feature}},  {\em Comptes Rendus
  Physique} {\bf 15} (Apr., 2014) 367--375,
  [\href{http://xxx.lanl.gov/abs/1403.5569}{{\tt arXiv:1403.5569}}].

\bibitem{linsley1963}
J.~{Linsley}, {\it {Primary cosmic rays of energy 10$^{17}$ to 10$^{20}$ eV,
  the energy spectrum and arrival directions}},  {\em Proceedings of the 8th
  International Cosmic Ray Conference} {\bf 4} (1963) 77.

\bibitem{wibig2005}
T.~Wibig and A.~W. Wolfendale, {\it {At what particle energy do extragalactic
  cosmic rays start to predominate?}},  {\em Journal of Physics G: Nuclear and
  Particle Physics} {\bf 31} (Mar., 2005) 255--264,
  [\href{http://xxx.lanl.gov/abs/astro-ph/0410624}{{\tt astro-ph/0410624}}].

\bibitem{allard2005}
D.~{Allard}, E.~{Parizot}, A.~V. {Olinto}, E.~{Khan}, and S.~{Goriely}, {\it
  {UHE nuclei propagation and the interpretation of the ankle in the cosmic-ray
  spectrum}},  {\em Astronomy \& Astrophysics} {\bf 443} (Dec., 2005) L29--L32,
  [\href{http://xxx.lanl.gov/abs/astro-ph/0505566}{{\tt astro-ph/0505566}}].

\bibitem{berezinsky2005}
V.~{Berezinsky}, A.~Z. {Gazizov}, and S.~I. {Grigorieva}, {\it {Dip in UHECR
  spectrum as signature of proton interaction with CMB}},  {\em Physics Letters
  B} {\bf 612} (Apr., 2005) 147--153,
  [\href{http://xxx.lanl.gov/abs/astro-ph/0502550}{{\tt astro-ph/0502550}}].

\bibitem{kascade2013b}
{KASCADE-Grande Collaboration}, {\it {KASCADE-Grande measurements of energy
  spectra for elemental groups of cosmic rays}},  {\em Astroparticle Physics}
  {\bf 47} (July, 2013) 54--66, [\href{http://xxx.lanl.gov/abs/1306.6283}{{\tt
  arXiv:1306.6283}}].

\bibitem{lemoine2005}
M.~{Lemoine}, {\it {Extragalactic magnetic fields and the second knee in the
  cosmic-ray spectrum}},  {\em Physical Review D} {\bf 71} (Apr., 2005) 083007,
  [\href{http://xxx.lanl.gov/abs/astro-ph/0411173}{{\tt astro-ph/0411173}}].

\bibitem{aloisio2005}
R.~Aloisio and V.~Berezinsky, {\it {Anti-GZK effect in ultra-high energy cosmic
  ray diffusive propagation}},  {\em The Astrophysical Journal} {\bf 625}
  (2005), no.~1 249--255.

\bibitem{kotera2008}
K.~Kotera and M.~Lemoine, {\it {Inhomogeneous extragalactic magnetic fields and
  the second knee in the cosmic ray spectrum}},  {\em Physical Review D} {\bf
  77} (Jan., 2008) 023005, [\href{http://xxx.lanl.gov/abs/0706.1891}{{\tt
  arXiv:0706.1891}}].

\bibitem{auger2008}
P.~A. Collaboration, {\it {Observation of the suppression of the flux of cosmic
  rays above 4$\times$10$^{19}$ eV}},  {\em Physical Review Letters} {\bf 101}
  (Aug., 2008) 061101, [\href{http://xxx.lanl.gov/abs/0806.4302}{{\tt
  arXiv:0806.4302}}].

\bibitem{hires2008}
H.~R. F.~E. Collaboration, {\it {First Observation of the
  Greisen-Zatsepin-Kuzmin Suppression}},  {\em Physical Review Letters} {\bf
  100} (Mar., 2008) 101101,
  [\href{http://xxx.lanl.gov/abs/astro-ph/0703099}{{\tt astro-ph/0703099}}].

\bibitem{greisen1966}
K.~Greisen, {\it {End of the Cosmic-Ray Spectrum?}},  {\em Physical Review
  Letters} {\bf 16} (1966), no.~17 748--750.

\bibitem{zatsepin1966}
G.~T. Zatsepin and V.~A. Kuz'min, {\it {Upper Limit of the Spectrum of Cosmic
  Rays}},  {\em Journal of Experimental and Theoretical Physics} {\bf 4}
  (1966), no.~3 114--117.

\bibitem{aloisio2011}
R.~Aloisio, V.~Berezinsky, and A.~Gazizov, {\it {Ultra high energy cosmic rays:
  The disappointing model}},  {\em Astroparticle Physics} {\bf 34} (Mar., 2011)
  620--626, [\href{http://xxx.lanl.gov/abs/0907.5194}{{\tt arXiv:0907.5194}}].

\bibitem{carilli2002}
C.~L. Carilli and G.~B. Taylor, {\it {Cluster Magnetic Fields}},  {\em Annual
  Review of Astronomy and Astrophysics} {\bf 40} (Sept., 2002) 319--348,
  [\href{http://xxx.lanl.gov/abs/astro-ph/0110655}{{\tt astro-ph/0110655}}].

\bibitem{ryu2011}
D.~{Ryu}, D.~R.~G. {Schleicher}, R.~A. {Treumann}, C.~G. {Tsagas}, and L.~M.
  {Widrow}, {\it {Magnetic Fields in the Large-Scale Structure of the
  Universe}},  {\em Space Science Reviews} {\bf 166} (May, 2012) 1--35,
  [\href{http://xxx.lanl.gov/abs/1109.4055}{{\tt arXiv:1109.4055}}].

\bibitem{ryu1998}
D.~{Ryu}, H.~{Kang}, and P.~L. {Biermann}, {\it {Cosmic magnetic fields in
  large scale filaments and sheets}},  {\em Astronomy \& Astrophysics} {\bf
  335} (July, 1998) 19--25,
  [\href{http://xxx.lanl.gov/abs/astro-ph/9803275}{{\tt astro-ph/9803275}}].

\bibitem{xu2006}
Y.~{Xu}, P.~P. {Kronberg}, S.~{Habib}, and Q.~W. {Dufton}, {\it {A Faraday
  Rotation Search for Magnetic Fields in Large-scale Structure}},  {\em The
  Astrophysical Journal} {\bf 637} (Jan., 2006) 19--26,
  [\href{http://xxx.lanl.gov/abs/astro-ph/0509826}{{\tt astro-ph/0509826}}].

\bibitem{neronov2010}
A.~{Neronov} and I.~{Vovk}, {\it {Evidence for Strong Extragalactic Magnetic
  Fields from Fermi Observations of TeV Blazars}},  {\em Science} {\bf 328}
  (Apr., 2010) 73--, [\href{http://xxx.lanl.gov/abs/1006.3504}{{\tt
  arXiv:1006.3504}}].

\bibitem{neronov2013}
A.~{Neronov}, A.~M. {Taylor}, C.~{Tchernin}, and I.~{Vovk}, {\it {Measuring the
  correlation length of intergalactic magnetic fields from observations of
  gamma-ray induced cascades}},  {\em Astronomy \& Astrophysics} {\bf 554}
  (June, 2013) A31, [\href{http://xxx.lanl.gov/abs/1307.2753}{{\tt
  arXiv:1307.2753}}].

\bibitem{syrovatskii1959}
S.~I. {Syrovatskii}, {\it {The Distribution of Relativistic Electrons in the
  Galaxy and the Spectrum of Synchrotron Radio Emission.}},  {\em Soviet
  Astronomy} {\bf 3} (Feb., 1959) 22.

\bibitem{berezinsky2006a}
V.~{Berezinsky} and A.~Z. {Gazizov}, {\it {Diffusion of Cosmic Rays in the
  Expanding Universe. I.}},  {\em The Astrophysical Journal} {\bf 643} (May,
  2006) 8--13, [\href{http://xxx.lanl.gov/abs/astro-ph/0512090}{{\tt
  astro-ph/0512090}}].

\bibitem{mollerach2013}
S.~{Mollerach} and E.~{Roulet}, {\it {Magnetic diffusion effects on the
  ultra-high energy cosmic ray spectrum and composition}},  {\em Journal of
  Cosmology and Astroparticle Physics} {\bf 10} (Oct., 2013) 13,
  [\href{http://xxx.lanl.gov/abs/1305.6519}{{\tt arXiv:1305.6519}}].

\bibitem{planck2013}
{Planck Collaboration}, {\it {Planck 2013 results. XVI. Cosmological
  parameters}},  {\em ArXiv e-prints} (Mar., 2013)
  [\href{http://xxx.lanl.gov/abs/1303.5076}{{\tt arXiv:1303.5076}}].

\bibitem{berezinsky1988}
V.~S. {Berezinskii} and S.~I. {Grigorieva}, {\it {A bump in the ultra-high
  energy cosmic ray spectrum}},  {\em Astronomy \& Astrophysics} {\bf 199}
  (June, 1988) 1--12.

\bibitem{aloisio2004}
R.~{Aloisio} and V.~{Berezinsky}, {\it {Diffusive Propagation of
  Ultra-High-Energy Cosmic Rays and the Propagation Theorem}},  {\em The
  Astrophysical Journal} {\bf 612} (Sept., 2004) 900--913,
  [\href{http://xxx.lanl.gov/abs/astro-ph/0403095}{{\tt astro-ph/0403095}}].

\bibitem{globus2008}
N.~{Globus}, D.~{Allard}, and E.~{Parizot}, {\it {Propagation of high-energy
  cosmic rays in extragalactic turbulent magnetic fields: resulting energy
  spectrum and composition}},  {\em Astronomy \& Astrophysics} {\bf 479} (Feb.,
  2008) 97--110, [\href{http://xxx.lanl.gov/abs/0709.1541}{{\tt
  arXiv:0709.1541}}].

\bibitem{miniati2002}
F.~{Miniati}, {\it {Intergalactic shock acceleration and the cosmic gamma-ray
  background}},  {\em Monthly Notices of the Royal Astronomical Society} {\bf
  337} (Nov., 2002) 199--208,
  [\href{http://xxx.lanl.gov/abs/astro-ph/0203014}{{\tt astro-ph/0203014}}].

\bibitem{dolag2005}
K.~{Dolag}, D.~{Grasso}, V.~{Springel}, and I.~{Tkachev}, {\it {Constrained
  simulations of the magnetic field in the local Universe and the propagation
  of ultrahigh energy cosmic rays}},  {\em Journal of Cosmology and
  Astroparticle Physics} {\bf 1} (Jan., 2005) 9,
  [\href{http://xxx.lanl.gov/abs/astro-ph/0410419}{{\tt astro-ph/0410419}}].

\bibitem{das2008}
S.~{Das}, H.~{Kang}, D.~{Ryu}, and J.~{Cho}, {\it {Propagation of
  Ultra-High-Energy Protons through the Magnetized Cosmic Web}},  {\em The
  Astrophysical Journal} {\bf 682} (July, 2008) 29--38,
  [\href{http://xxx.lanl.gov/abs/0801.0371}{{\tt arXiv:0801.0371}}].

\bibitem{donnert2009}
J.~{Donnert}, K.~{Dolag}, H.~{Lesch}, and E.~{M{\"u}ller}, {\it {Cluster
  magnetic fields from galactic outflows}},  {\em Monthly Notices of the Royal
  Astronomical Society} {\bf 392} (Jan., 2009) 1008--1021,
  [\href{http://xxx.lanl.gov/abs/0808.0919}{{\tt arXiv:0808.0919}}].

\bibitem{auger2013a}
{Pierre Auger Collaboration}, {\it {Bounds on the density of sources of
  ultra-high energy cosmic rays from the Pierre Auger Observatory}},  {\em
  Journal of Cosmology and Astroparticle Physics} {\bf 5} (May, 2013) 9,
  [\href{http://xxx.lanl.gov/abs/1305.1576}{{\tt arXiv:1305.1576}}].

\bibitem{sigl2007}
G.~Sigl, {\it {Nonuniversal spectra of ultrahigh energy cosmic ray primaries
  and secondaries in a structured universe}},  {\em Physical Review D} {\bf 75}
  (May, 2007) 103001, [\href{http://xxx.lanl.gov/abs/astro-ph/0703403}{{\tt
  astro-ph/0703403}}].

\bibitem{aloisio2009}
R.~Aloisio, V.~Berezinsky, and A.~Gazizov, {\it {The Problem of Superluminal
  Diffusion of Relativistic Particles and Its Phenomenological Solution}},
  {\em The Astrophysical Journal} {\bf 693} (Mar., 2009) 1275--1282.

\bibitem{hooper2010}
D.~Hooper and A.~M. Taylor, {\it {On the heavy chemical composition of the
  ultra-high energy cosmic rays}},  {\em Astroparticle Physics} {\bf 33} (Apr.,
  2010) 151--159, [\href{http://xxx.lanl.gov/abs/0910.1842}{{\tt
  arXiv:0910.1842}}].

\bibitem{taylor2011}
A.~M. Taylor, M.~Ahlers, and F.~A. Aharonian, {\it {Need for a local source of
  ultrahigh-energy cosmic-ray nuclei}},  {\em Physical Review D} {\bf 84}
  (Nov., 2011) 105007, [\href{http://xxx.lanl.gov/abs/1107.2055}{{\tt
  arXiv:1107.2055}}].

\bibitem{allard2012}
D.~Allard, {\it {Extragalactic propagation of ultrahigh energy cosmic-rays}},
  {\em Astroparticle Physics} {\bf 39-40} (Dec., 2012) 33--43,
  [\href{http://xxx.lanl.gov/abs/1111.3290}{{\tt 1111.3290}}].

\bibitem{aloisio2013b}
R.~{Aloisio}, V.~{Berezinsky}, and P.~{Blasi}, {\it {Ultra high energy cosmic
  rays: implications of Auger data for source spectra and chemical
  composition}},  {\em ArXiv e-prints} (Dec., 2013)
  [\href{http://xxx.lanl.gov/abs/1312.7459}{{\tt arXiv:1312.7459}}].

\bibitem{arons2003}
J.~Arons, {\it {Magnetars in the Metagalaxy: An Origin for UltraÐHigh?Energy
  Cosmic Rays in the Nearby Universe}},  {\em The Astrophysical Journal} {\bf
  589} (June, 2003) 871--892,
  [\href{http://xxx.lanl.gov/abs/astro-ph/0208444}{{\tt astro-ph/0208444}}].

\bibitem{fang2012}
K.~Fang, K.~Kotera, and A.~V. Olinto, {\it {Newly Born Pulsars As Sources of
  Ultrahigh Energy Cosmic Rays}},  {\em The Astrophysical Journal} {\bf 750}
  (May, 2012) 118, [\href{http://xxx.lanl.gov/abs/1201.5197}{{\tt
  arXiv:1201.5197}}].

\bibitem{fang2013}
K.~Fang, K.~Kotera, and A.~V. Olinto, {\it {Ultrahigh energy cosmic ray nuclei
  from extragalactic pulsars and the effect of their Galactic counterparts}},
  {\em Journal of Cosmology and Astroparticle Physics} {\bf 2013} (Mar., 2013)
  010--010, [\href{http://xxx.lanl.gov/abs/1302.4482}{{\tt arXiv:1302.4482}}].

\end{thebibliography}\endgroup
\bibliographystyle{jhep}

\end{document}